\documentclass[aps,pra,twocolumn,noshowpacs,superscriptaddress,groupedaddress]{revtex4}  
\usepackage{graphicx}  
\usepackage{dcolumn}   
\usepackage{bm}        
\usepackage{amssymb}   
\usepackage{amsmath}
\usepackage{color}
\usepackage[free-standing-units=true]{siunitx} 
\usepackage[colorlinks=true, pdfstartview=FitV, linkcolor=blue, citecolor=blue, urlcolor=blue]{hyperref} 
\usepackage{epstopdf}

\usepackage{appendix} 

\newcommand{\affilITP}{Institute for Theoretical Physics, ETH Z\"{u}rich, CH-8093 Z\"urich, Switzerland.}
\newcommand{\affilIFP}{Laboratory for Solid State Physics, ETH Z\"{u}rich, CH-8093 Z\"urich, Switzerland.}
\newcommand{\affilQC}{Quantum Center, ETH Zurich, CH-8093 Zurich, Switzerland}
\newcommand{\affilUKON}{Department of Physics, University of Konstanz, D-78457 Konstanz, Germany}

\begin{document}

\preprint{}

\title{Proliferation of unstable states \\and their impact on stochastic out-of-equilibrium dynamics}

\author{Toni L. Heugel}\affiliation{\affilITP}
\author{R. Chitra}\affiliation{\affilITP}
\author{Alexander Eichler}\affiliation{\affilIFP}\affiliation{\affilQC}
\author{Oded Zilberberg}\affiliation{\affilUKON}
%
%
\date{\today}


\begin{abstract}
Networks of nonlinear parametric resonators are promising candidates as Ising machines for annealing and optimization. 
These many-body out-of-equilibrium systems host complex phase diagrams of  coexisting stationary states. The plethora of states manifest via a series of bifurcations, including bifurcations that proliferate purely unstable solutions, which  we term ``ghost bifurcations''. Here, we demonstrate that the latter take a fundamental role in the stocahstic dynamics of the system in the presence of noise. Specifically, they determine the switching paths and the switching rates between stable solutions.  We demonstrate experimentally the impact of ghost bifurcations on the noise-activated switching dynamics in a network of two coupled parametric resonators. 
\end{abstract}
\maketitle





\section{Introduction}
Statistical physics has provided valuable insights into the phenomenon of noise-induced switching between local energy minima, as exemplified by the well-known Kramers double-well problem~\cite{Kramers1940, Hanggi_1990, rondin2017direct}.
Such stochastic dynamics is highly relevant for a wide variety of phenomena spanning protein folding~\cite{Best2006,Chung2015}, chemical reactions~\cite{GarciaMuller2008}, as well as stability in mechanical~\cite{Badzey2005,Rondin2017} and electrical systems~\cite{Fulton1974,Silvestrini1988}. While the noisy dynamics of equilibrium  systems have been extensively studied, that of systems driven far out of equilibrium remains largely unexplored~\cite{Dykman_1998,Dykman_2007,tadokoro2020noise}. 

An important class of out-of-equilibrium systems are driven systems. Such systems are characterized by stationary oscillation states that manifest when the conserving and nonconserving forces in the system are in balance. For nonlinear systems, there may be several such stationary states that act as attractors, just like potential wells do in equilibrium systems. In a rotating frame, the resulting dynamics can resemble that of an equilibrium potential landscape~\cite{Hanggi_1990}. Extending the analogy, noise can induce stochastic switching between the attractors, and the switching rate can be treated with the abstract notion of a potential activation barrier~\cite{Dykman_1998,Luchinsky1999,lapidus,kkim2005,Aldridge2005,Chan_2007,Chan_2008,Venstra2013,Mahboob_2014_2,margiani2022extracting}. However, it is important to emphasize that this activation barrier is not related to a gap in free energy, but instead to a ``phase gap'' that separates the attractors~\cite{Frimmer_2019}. Understanding how often the system switches between attractors, and which path it selects during the switch, requires different methods and can be cumbersome~\cite{Hanggi_1990,Dykman_1998}.

A paradigmatic example of a bistable out-of-equilibrium system is the Kerr parametric oscillator (KPO)~\cite{Ryvkine_2006, Mahboob_2008, Wilson_2010, Eichler_2011_NL, Leuch_2016, Gieseler_2012, Lin_2014, Puri_2017, Eichler_2018, Nosan_2019, Frimmer_2019, Grimm_2019, Puri_2019_PRX, Miller_2019_phase, DykmanBook,Eichler_Zilberberg_book}.
Recently, networks of coupled, driven KPOs have been proposed as a simulation platform to solve complex problems optimally.~\cite{ Mahboob_2016, Inagaki_2016, Goto_2016, Puri_2017_NC,Nigg_2017,Dykman_2018,Okawachi_2020}. Such networks typically possess a large number of stationary states, analogous to an multi-well potential with rich phase transitions~\cite{Strinanti_2021,Bello_2019,Heugel_2022,heugel2022role}. Studying activated switching between states in the presence of fluctuations is crucial for understanding their stability and lifetimes. It will influence how these networks are operated, and it can also provide a characterization method that is unaffected by the danger of local trapping~\cite{margiani2022deterministic}.

In this work, we address the physics of stochastic activation in a  tractable system of two strongly coupled, classical KPOs. The system possesses various stable and unstable stationary states~\cite{Heugel_2019_TC,Heugel_2022}, and we observe stochastic switching between two such states in the presence of fluctuations. Surprisingly, the switching rate $\Gamma$ deviates significantly from the exponential model expected for a single KPO~\cite{Dykman_1998}. Seeking to explain this deviation, we calculate the dominant transition paths between the states with the Onsager-Machlup function~\cite{Lehmann_2003, Wio_2013}. Our analysis shows that new kinds of unstable states, termed ghost states  emerge in the two KPO system. These states do not manifest in the stationary deterministic dynamics of the system. Interestingly, however, they offer new transition paths and contribute significantly to the transition paths and the corresponding transition rates. We thus identify a striking example of out-of-equilibrium statistical physics in a nonlinear multistable system. Our work paves the way for the exploration of larger systems, especially in view of KPO networks as solvers for complex optimization tasks.

\section{System}  
In the following, we analyze noise-induced switching dynamics using an experimental setup composed of two electrical KPOs with capacitive coupling~\cite{Nosan_2019,Heugel_2022}. Each KPO consists of a coil with inductance $L$ and a diode that provides a nonlinear capacitance $C$, cf. Appendix~\ref{sec_single_KPO}. The resonance frequency of each KPO can be tuned by applying a DC voltage across the diode. 
We drive and measure the resonators inductively through auxiliary coils. Our electrical circuits are well described by the following coupled equations of motion
\begin{multline}
	\ddot{x}_i + \omega_i^2\left[1-\lambda\cos\left(2\omega_d t\right)\right]x_i + \alpha_i x_i^3 \\+ \gamma_i \dot{x}_i -\sum_{j\neq i}J_{ij} x_j = \xi_i(t)\,, \label{eq:coupled_EOM}
\end{multline}
where dots indicate time derivatives, $x_i = u_i \cos(\omega t) - v_i \sin(\omega t)$ is the measured voltage with quadrature amplitudes $u_i$ and $v_i$, $\omega_i = 2\pi f_i$ is the angular eigenfrequency, and $J_{ij}$ ($i\neq j$) denotes the inter-circuit linear coupling strength. Each resonator has an effective Duffing (Kerr) nonlinearity with coefficient $\alpha_i$ and a damping rate $\gamma_i = \omega_i/Q_i$, with  $Q_i$ the quality factor. Our resonators are constructed and tuned to be (nearly) identical in their bare characteristics, $\omega_i \approx \omega_0 = 2\pi f_0$.  The same parametric pumping with angular modulation frequency $2\omega_d = 4\pi f_d \approx 2\omega_0$ and modulation depth $\lambda \propto U_d$ is applied to all resonators. 

Crucially, beyond a frequency-dependent driving threshold $U_\mathrm{th}$, the KPO has exactly two stable solutions that we refer to as `phase states', which have identical amplitudes but differ in phase by $\pi$~\cite{mclachlan1951theory,Lifshitz_Cross,Eichler_Zilberberg_book}. These are the two attractors of a single KPO in a frame rotating at $\omega_d$. In order to induce switching events between the attractors, we add an artificial noise $\xi_i$ generated by a fluctuating voltage  with white power spectral density $S_n$. The noise $\xi_i$ simulates a thermal force noise with $\langle \xi_i(t_1) \xi_j(t_2)\rangle = \varsigma^2 \delta_{ij} \delta(t_1-t_2)$ and  power spectral density calibrated to be $\varsigma^2 = \SI{4.93e-20}{\hertz^4} S_{n}$, see Appendix~\ref{sec:fluctuating_coherent}.

In our experiments, we use a lock-in amplifier to measure the quadratures $(u_i, v_i)$ which vary on timescales much longer than $1/\omega_0$. The evolution of these quadratures is well
captured in the rotating-frame picture obtained by applying the averaging method~\cite{guckenheimer_1990, Papariello_2016, Heugel_2019_TC}. to our model. Equation~\eqref{eq:coupled_EOM} then leads to the following slow-flow equations: 
\begin{align}
    \label{eq:slowflow}
    \dot{u}_i &= -\frac{\gamma  u_i}{2}- \left(\frac{3 \alpha}{8 \omega_d }X_i^2 +\frac{\omega_0^2 - \omega_d^2}{2\omega_d}+\frac{\lambda \omega_0^2}{4 \omega_d }\right) v_i +\frac{J  v_j}{2 \omega_d }+\Xi_{u_i}\,,\nonumber\\
    \dot{v}_i &= -\frac{\gamma  v_i}{2} + \left(\frac{3 \alpha}{8 \omega_d }X_i^2 + \frac{\omega_0^2 - \omega_d^2 }{2\omega_d} -\frac{\lambda \omega_0^2}{4 \omega_d }\right)u_i-\frac{J  u_j}{2 \omega_d }+\Xi_{v_i}\,,
\end{align}
where $X_i^2 = u_i^2 + v_i^2$, and we have additive uncorrelated noise terms $\Xi_{u_i}$, $\Xi_{v_i}$, whose power spectral densities are given by $\sigma^2=~ \varsigma^2/2\omega_d^2$~\cite{Khasminskii_66, Roberts_86}. As shown in previous works, the averaging method fully captures the physics of our
networks in the regime where $\lambda$, $\gamma/\omega_0$, $J/\omega_0^2$, $(\alpha/\omega_0^2)x_i^2$, and $\alpha \varsigma^2 / \omega_0^5$ are all of order $\epsilon$ with $0<\epsilon\ll 1$~\cite{nayfeh2008,Khasminskii_66,Papariello_2016,Eichler_2018}. Here and in the following, we assume identical dissipation rates $\gamma_i = \gamma$, nonlinearities $\alpha_i = \alpha$, and coupling $J_{ij}=J$.

\begin{figure}[t!]
    \includegraphics[width=86mm]{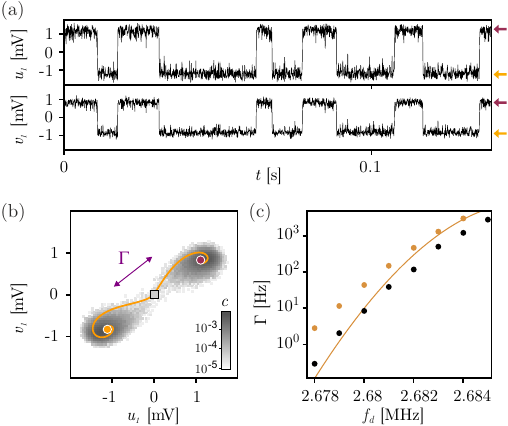}
    \caption{Noise-induced switching in a single KPO. (a)~Time trace of the rotating quadratures $u_1$ and $v_1$, showing noisy fluctuations around and switching between the two phase states for $U_d=\SI{3}{\volt}$ and $S_{n} = \SI{2.75e-8}{\volt\squared\per\hertz}$ at $f_d = \SI{2.681}{\mega\hertz}$. Arrows indicate the position of the phase states. (b)~Heatmap of normalized measurement counts $c$ versus $u_1$ and $v_1$, showing two dominant attractors around the phase states (colored dots). The data stem from a time trace measurement as in (a) taken over \SI{8}{\second}. Counts close to the origin arise due to inter-state switching with rate $\Gamma$. The orange line shows the most probable switching path through the origin (grey square) predicted from the Onsager-Machlup formalism, cf. Eq.~\eqref{eq:Onsager_Machlup} and Appendix~\ref{sec_single_KPO}. (c)~Switching rate $\Gamma$ as a function of $f_d$ with the same parameters as in (a) obtained from the experiment (black dots)~\cite{margiani2022extracting}. The orange line corresponds to the approximate analytical result from Ref.~\cite{Dykman_1998}, cf. Appendix~\ref{sec:gamma_det}, and the orange dots to numerical minimization of Eq.~\eqref{eq:Onsager_Machlup}.}
    \label{fig:Fig_1}
\end{figure}



\section{Transition Rates}
A switching experiment with KPO 1 (while KPO 2 is detuned) is shown in Fig.~\ref{fig:Fig_1}(a). There, we observe that the system resides in each phase state for a certain dwell time before switching to the opposite state. The average dwell time $\tau$ can be expressed as a rate of activated switching $\Gamma = \tau^{-1}$. In the rotating phase space, the same measurement data can be represented as a density of count rates, see Fig.~\ref{fig:Fig_1}(b). As discussed in Ref.~\cite{Dykman_1998}, the logarithm of the switching rate is inversely proportional to the distance between the two phase states. In Fig.~\ref{fig:Fig_1}(c), this leads to an exponential decrease in $\Gamma$ with decreasing $f_d$ (the direction in $f_f$ in which $\Gamma$ increases depends on the sign of the nonlinearity).

\begin{figure*}[t!]
  \includegraphics[width=178mm]{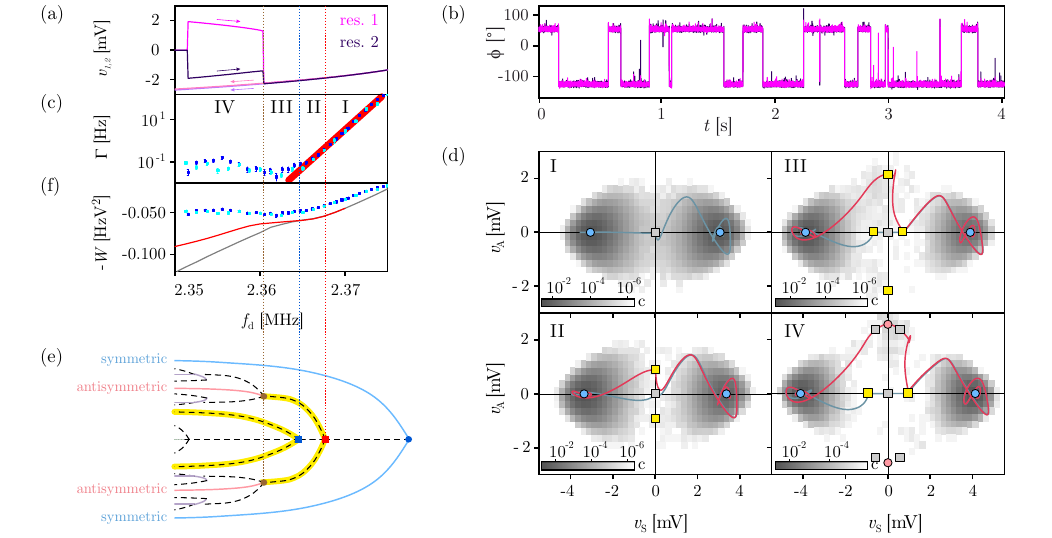}
  \caption{Noise-induced switching in the two-KPO system.
  (a)~Quasi-static frequency sweep with a parametric drive $U_d=\SI{3.7}{\volt}$ applied to both resonators. The full data (including the $u_{1,2}$ and the stability diagram) are shown in Fig.~\ref{fig:FigS1}. Arrows indicate the sweep direction. Vertical dashed lines indicate the bifurcation frequencies, c.f. (e). (b)~Switching between the two symmetric phase states at $f_d = \SI{2.37}{\mega\hertz}$ in the presence of applied voltage noise (see Appendix~\ref{sec:fluctuating_coherent} for details).
  (c)~Switching rate $\Gamma$ as a function of $f_d$ for a noise PSD $S_n = \SI{1.1e-7}{\volt\squared\per\hertz}$. Dark (light) blue corresponds to experimental (simulated) data with error bars estimated from textbook Poisson statistics. A thick red line reflects the exponential trend.
  (d)~Switching between the two symmetric states visualized in the rotating frame in terms of $v_S$ and $v_A$. I: \SI{2.37}{\mega\hertz}, II: \SI{2.3675}{\mega\hertz}, III: \SI{2.3625}{\mega\hertz}, IV: \SI{2.36}{\mega\hertz}. Greyscale heat maps represent the normalized counts $c$ from multiple switches. Circles and squares indicate stable and unstable solutions, respectively. The ghost states are marked in yellow. The solid lines are the theoretically predicted switching paths between the stable symmetric states using the Onsager-Machlup formalism, cf. Eq.~\eqref{eq:Onsager_Machlup}. The color indicates whether a path passes the unstable ghost state (red) or the unstable 0-amplitude state (blue-gray). 
  (e)~Schematics of the steady states and bifurcation points as a function of $f_d$. Solid and dashed lines are for stable  and unstable solutions, respectively. Squares mark ghost bifurcations that proliferate additional unstable states (ghost states) from an already unstable attractor. Ghost states are marked in yellow.
  (f)~Activation $W$ as a function of $f_d$. Dark (light) blue corresponds to experimental (simulated) data with error bars estimated from textbook Poisson statistics. The solid gray and red lines are obtained by minimizing $S_{\rm OM}$ for paths via the 0-amplitude state and the antisymmetric state, respectively.
  The system parameters are: $Q = 265$, $f_0 = \SI{2.3670}{\mega\hertz}$, $\alpha = \SI{-6.5e17}{\per\square\volt\per\square\second}$, $U_{th} = \SI{1.73}{\volt}$, and $J = \SI{-1.28}{\mega\hertz^2}$.}
  \label{fig:Fig_2}
\end{figure*}

We now consider the case that the two KPO are tuned to have the same frequency. The experimental phase diagram of the system as a function of $\lambda$ and $\omega_p$ was characterized in Ref.~\cite{Heugel_2022}. 
In Fig.~\ref{fig:Fig_2}(a), we show the results of a measured frequency sweep passing through various regions containing up to four different types of stable two-oscillator states: a state with both resonators having amplitude zero; (S) symmetric states, i.e., the two oscillators have the same phase; (A) antisymmetric states where the two oscillators have $\pi$-shifted phases; and (M) mixed-symmetry states that are neither symmetric nor antisymmetric. The symmetric/antisymmetric solutions can be interpreted as the parametrically driven symmetric/antisymmetric normal modes of the two resonators. See Fig.~\ref{fig:FigS1} for more information regarding the stability diagram.

Typically, when initialized in one such state, the resonator explores the vicinity of the attractor under the influence of the noise terms $\Xi_{u_i}$ and $\Xi_{v_i}$ ~\cite{heugel2022role}. Occasionally, the noise activates the system over the quasi-potential barrier to reach a different attractor, as seen for instance in Fig.~\ref{fig:Fig_2}(b)~\cite{margiani2022extracting}. The system can switch back and forth in time and is characterized by a switching rate $\Gamma$. In the following, we analyze $\Gamma$ along the sweep in Fig.~\ref{fig:Fig_2}(a) for $f_d>\SI{2.36}{\mega\hertz}$. In this frequency range, only the symmetric solution is stable~\cite{Heugel_2022}. We therefore expect $\Gamma(f_d)$ to manifest an exponential behaviour analogous to that seen in a single KPO in Fig.~\ref{fig:Fig_1}. Motivated by this idea, we set out to confirm the hypothesis that each normal mode has the same activation rate scaling as a single KPO.


In Fig.~\ref{fig:Fig_2}(c), we show the measured transition rate $\Gamma$ for switches between the two symmetric states as a function of $f_d$ while the parametric driving strength $U_d$ is fixed. With decreasing $f_d$, the transition rate $\Gamma$ decreases exponentially as expected from previous single KPO studies~\cite{Dykman_1998, Ryvkine_2006, Stambaugh_2006, Chan_2008,margiani2022extracting}. This holds for a distinct range marked as I and II until $f_d \approx \SI{2.365}{\mega \hertz}$ is reached. Surprisingly, below this frequency, we observe substantial deviations from the simple exponential model even though the symmetric state remains the only stable configuration. First, a kink  manifests at $f_d= \SI{2.365}{\mega\hertz}$, implying a relative enhancement of the switching in region III. At even lower frequencies, the slope of $\Gamma$ changes sign, signaling a significant increase of the switching with decreasing frequency in region IV. 
This behaviour is fundamentally at odds with the
standard expectation of  decreasing rates.

To obtain deeper insights into the curious switching behaviour of the two-KPO system, we look at several measured transition events and systematically collect the 4-quadrature state vector $\mathbf{Y}=(u_1,v_1,u_2,v_2)$, cf. Appendix~\ref{sec:antisymm_switching}.
To visualize the transitions in a 2-dimensional space, we use symmetric and antisymmetric coordinates, $v_{\rm S} = (v_1 + v_2)/\sqrt{2}$,  $v_{\rm A} = (v_1 - v_2)/\sqrt{2}$ (and analogous for $u$).
By plotting several time traces containing multiple transition events in the phase space spanned by $v_S$ and $v_A$, we obtain the corresponding probability distribution. 

Comparing the distributions at four representative frequencies, we find striking differences, see Fig.~\ref{fig:Fig_2}(d). Regions I and II are characterized by two attractors with high probabilities, corresponding to the two phase states of the symmetric mode. By contrast, for $f_d < \SI{2.365}{\mega\hertz}$ in the regions III and IV, we find the emergence of a substantial probability centered around the attractors of the antisymmetric modes. This indicates that the appearance of the kink at $f_d=\SI{2.365}{\mega\hertz}$ follows a drastic change of the transition path chosen by the system, and of the underlying quasi-potential landscape. Such a deviation in region IV is to be expected as the activation dynamics now involves four different stable attractors. In other words, the naive model of activation between two states is insufficient in region IV. This is confirmed by a numerical simulation of the noisy time evolution of the EOMs, given in Eq.~\eqref{eq:slowflow}, which is in accord with the experimental results, see bright blue data in Fig.~\ref{fig:Fig_2}(c). Crucially, however, we note that the antisymmetric state is not stable in region III, as indicated by a red square (instead of a circle). This observation raises the question: why should the unstable solutions of the syste influence the transition paths?
In the following, we precisely address this question through an in-depth theoretical study of the transition dynamics.

\section{Transition Paths}
The weak noise-induced switching between stable oscillation states is analogous to noise-activated jumping over a barrier $W$  studied in an equilibrium system~\cite{Stambaugh_2006,Hanggi_1990}. The principal difference is that in our driven system,  the barrier $W$ between two stable attractors resides in a quasipotential structure in a rotating frame. We use the  Onsager-Machlup formalism to identify the optimal transition paths in phase space between two stable attractors, whose corresponding action then provides an estimation for the barrier $W$. We first define the Onsager-Machlup function~\cite{Lehmann_2003, Wio_2013} 
\begin{equation}\label{eq:Onsager_Machlup}
    S_{\rm OM}[\mathbf{Y}] = \int_{t_i}^{t_f} \frac{1}{4}\left(\dot{\mathbf{Y}}-\mathbf{f}(\mathbf{Y}) \right)^2 dt\,,
\end{equation}
where $t_i$ ($t_f$) is the initial (final) time of the trajectory of a system composed of $N$ resonators, $\mathbf{Y}=(u_1,v_1,...,u_N,v_N)^T$, and $\mathbf{f}(\mathbf{Y})$ is the right hand side of Eq.~\eqref{eq:slowflow} without noise terms written as a column vector.
The switching probability density between two stable states  connected by a  path $\mathbf{Y}(t)$ is given by $e^{-2 S_{\rm OM}[\mathbf{Y}]/\sigma^2}$. The total switching probability $P_{if}$ from  a stable attractor at $\mathbf{Y}_i$ to one at $\mathbf{Y}_f$ is obtained by integration over all  allowed trajectories connecting them. From this probability one can derive the switching rate $\Gamma$~\cite{Lehmann_2003}, which in the weak-noise limit scales as $\Gamma \propto \exp(-2 W/\sigma^2)$ with barrier $W$~\cite{Stambaugh_2006}.


%


At low noise, the switching rate $\Gamma$ is dominated by the path $\mathbf{Y}_{\rm min}$ that minimizes $S_{\rm OM}$~\cite{Lehmann_2003}. Hence, we can neglect the integration over all possible paths and the switching rate is approximately given by
\begin{equation}
\label{eq:Gamma}
\Gamma \approx \Gamma_{\rm min}\equiv \Gamma_0 e^{-2 S_{\rm OM}[\mathbf{Y}_{\rm min}]/\sigma^2}\,, 
\end{equation}
where $\Gamma_0$ is an overall prefactor and we identify the effective activation barrier $W_{\rm eff}=S_{\rm OM}[\mathbf{Y}_{\rm min}]$~\cite{Lehmann_2003}. If $S_{\rm OM}$ has multiple local minima, one needs to find the contribution from all relevant minimizing paths and weigh their relative contributions.




The landscape of $S_{OM}$ is characterized by a bifurcation diagram of the system equations, which categorize both stable and unstable solutions as effective minima and saddles/maxima, respectively. In Fig.~\ref{fig:Fig_2}(d), we schematically show this bifurcation diagram along the frequency sweep in Fig.~\ref{fig:Fig_2}(a). Various pitchfork bifurcations lead to the emergence of stable and unstable states. Interestingly, some of the bifurcations involve purely unstable states and are not visible in a frequency sweep. We therefore dub them ``ghost bifurcations'' as they leave the deterministic steady state physics unaffected. Such ghost bifurcations separate regions I and II, as well as II and III. At each ghost bifurcation, additional unstable states emerge. To emphasize their origin, we refer to these unstable states as ``ghost states''. From comparing Fig.~\ref{fig:Fig_2}(c) and (e), it becomes clear that the emergence of the ghost states is accompanied by new switching paths, and that these changes impact $\Gamma$. Specifically, rather than acting as additional obstacles in the switching paths, the ghost states appear to favor an \textit{increase} in $\Gamma$.

To elucidate the surprising role of the ghost states, we study the transition paths $\mathbf{Y}_{\rm min}$ and the corresponding barriers $W$ for the different representative frequencies along the sweep shown in Fig.~\ref{fig:Fig_2}(a). To this end, we need to minimize  Eq.~\eqref{eq:Onsager_Machlup} which is a complex task. A simple variational scheme with equal timesteps gives inconsistent results and more advanced methods as the sgMAM-method~\cite{Grafke_2017} are necessary  to obtain the correct physical paths (see details in Appendix~\ref{sec:gamma_det}). 

In Fig.~\ref{fig:Fig_2}(d), we show the calculated switching paths corresponding to the experimental parameters, which are representative for the regions (I-IV). In region I where the symmetric states are the only stable attractors, we find only one switching path, which passes through the intermediate 0-amplitude state in agreement with our experimentally observed distributions. As expected, this is 
equivalent to the single KPO case in Fig.~\ref{fig:Fig_1}, confirming that we should expect a monotonic decrease transition rate $\Gamma(f_d)$ with decreasing frequency as the stable attractors move apart~\cite{Dykman_1998,Chan_2008}.  

In region II, we find two additional switching paths that avoid the 0-amplitude state and instead pass through emergent unstable antisymmetric states. This is in line with the experimental data that exhibits a broader distribution around the 0-amplitude state, extending to the two unstable antisymmetric states. Similar paths also arise in regions III and IV, where the unstable states provide transient ledges where the system can hover during switching events, cf. Appendix~\ref{sec:antisymm_switching}. The additional switching paths are visible in the experiment and seem to be the dominant paths in regions III and IV. These alternative switching paths  bring forth a complexity to the system dynamics, a feature intrinsically related to the existence of multiple normal modes in KPO networks. In our system, we observe that the minimal path, $\mathbf{Y}_{\rm min}$, connects the stable states always via an unstable one in see Fig.~\ref{fig:Fig_2}(d)~\cite{Maier_1992, Tang_2017}, although exceptions have been observed~\cite{Luchinsky_1999, Feng_2014}. This underscores the importance of investigating stable as well as unstable states of the system in order to understand the stochastic dynamics of Ising networks. 

Based on our theoretical analysis of optimal switching paths, we can now obtain $W$ and compare it with that extracted from the experimental data, cf. Appendix~\ref{sec:gamma_det}. From this, we identify the dominant transition path, see Fig.~\ref{fig:Fig_2}(d).
Within regions I-III, we find good qualitative and approximate quantitative agreement (save for a small overall shift). Interestingly, in region II, both the paths contribute equally to the activation: one via the antisymmetric ghost states and the other via the origin. However, this effect is not sufficiently strong to be observed in Fig.~\ref{fig:Fig_2}(c). In region III, the former overtakes the latter, which manifests as the observed kink in Fig.~\ref{fig:Fig_2}(c). This indicates that the ghost states support the antisymmetric switching, and markedly participate and modify the expected stochastic dynamics of the system. In region IV, both symmetric and antisymmetric phase states are stable. Here, the analytical calculation deviates from the experimental and numerical results. This deviation is likely due to the fact that the Onsager-Machlup method only considers switches that connect the two symmetrical states, while the counting algorithm that we used for the experimental and numerical data includes all possible switches between states. The latter includes repeated switches between a symmetric and an antisymmetric state as individual events.


\section{Discussion}
At a fundamental level, our results demonstrate unambiguously the existence of an unusual type of bifurcation arising from the coupling between the individual resonators. Although these ghost bifurcations remain undetected in a deterministic system characterization~\cite{Heugel_2022}, they impact the inter-state switching path, switching rate $\Gamma$, and the relative dwell times in the symmetric and antisymmetric states. This inevitably affects the stochastic switching processes in KPO networks, and their characterization via stochastic sampling~\cite{margiani2022deterministic}.

Generalizing our results to $N$ identical resonators with identical all-to-all coupling, we theoretically predict that the ghost bifurcations, as well as the altered stability of the  symmetric/antisymmetric regions, persist. Such a network has a single symmetric lobe and a $N-1$-fold degenerate antisymmetric lobe~\cite{Heugel_2019_TC}. The boundaries delineating the overlapping region once more involve ghost bifurcations. However, one of the ghost bifurcations is now $N-1$ fold degenerate. As demonstrated above for two resonators, the antisymmetric states only become stable when they undergo another bifurcation, described by the general equation
\begin{equation}
\lambda_A = \frac{\sqrt{4 \gamma ^2 \omega ^2+\left(J (N+2)-2 \left(\omega ^2-\omega_0^2\right)\right)^2}}{\omega_0^2}\, ,
\end{equation}
resulting in an extended symmetric region. We thus expect a KPO network with identical all-to-all coupling to manifest the four different regimes of activation seen in the two-KPO case. When the coupling coefficient $J_{ij}$ is different for each resonator pair, the degeneracy is lifted. However, the instability lobes still overlap and generally form ghost bifurcations which will also impact the switching behaviour of the system.  In analogy to our observations in Fig.~\ref{fig:Fig_2}, many competing switching paths open up. Finding the dominant switching paths is a complex and demanding task. Our work provides the motivation for such future investigations.

All of our observations have important consequences for logic networks built from KPOs and nonlinear resonators in general, because they impact the solution that a network will choose after a finite transition time. 
With proper modelling and calibration of the ghost bifurcations, a network can be operated at a position in $f_d-\lambda$ space where the many-body character of the network is preserved and the annealing speed is optimized.
It becomes clear from our work that large networks bear very complex switching dynamics and a careful analysis of the bifurcation topology is very important. Future work might find ways to use this complexity in an advantageous manner to perform faster calculations.

\section{Summary and Outlook} 
In summary, we experimentally and theoretically investigated the noise-induced dynamics of a system of two coupled nonlinear Kerr parametric oscillators (KPOs). Our study implements the smallest form of a KPO network, and tests the switching behavior in so-called Ising machines. 
We found that ghost bifurcations play an important role, with consequences for the switching dynamics of the system as it progresses towards its most stable configuration. A better understanding of such effects can be very helpful for the calibration to and for stochastic logic protocols, such as simulated annealing. As coupled networks of parametric resonators are one of the main candidates for future parallel computation architectures, our study provides crucial input for a growing subcommunity working towards classical and quantum analog computation~\cite{Gottesman_2001, Devoret_2013, Mahboob_2016, Inagaki_2016, Goto_2016, Puri_2019_PRX,Nigg_2017,Dykman_2018}. Furthermore, it provides additional incentives for the fundamental exploration of complex driven-dissipative nonlinear networks in a multitude of fields~\cite{DykmanBook}.


\acknowledgments
This work received financial support from the Swiss National Science Foundation through grants (CRSII5\_206008/1) and (PP00P2\_163818), and the Deutsche Forschungsgemeinschaft (DFG) through project number 449653034 and SFB1432. We thank Peter M\"{a}rki, \v{Z}iga Nosan and Christian Marty for technical help.



\appendix

\section{Single KPO}\label{sec_single_KPO} 

\begin{figure}[t!]
    \includegraphics[width=\linewidth]{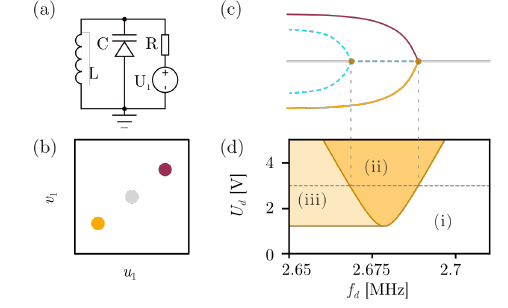}
    \caption{(a)~Schematics of a parametric RLC resonator  with resistance $R$, inductance $L$, nonlinear capacitance $C$, and tuning voltage $U_{1}$. 
  (b)~Schematic representation of the parametric phase states (wine red, orange), and the 0-amplitude state (grey) in phase space. (c)~Schematics of the steady states and bifurcation points as a function of $f_d$. Solid (dashed) lines correspond to stable (unstable) solutions. 
  (d)~Stability phase diagram: (i) White: only the 0-amplitude solution is stable; (ii) orange: only the phase states are stable; (iii) light orange: 0-amplitude and phase states are both stable.}
    \label{fig:Fig_Single_Steady}
\end{figure}

For $J=0$, each resonator can be driven into parametric resonance when $U_d \geq U_{th}$~\cite{Landau_Lifshitz, Lifshitz_Cross}. We characterize each resonator using frequency sweeps as described in~\cite{Leuch_2016,Heugel_2019_TC,Nosan_2019} and obtain the values $Q_1 = 295$, $f_0 = \SI{2.6784}{\mega\hertz}$, $\alpha_1 = \SI{-9e17}{\per\square\volt\per\square\second}$, and $U_{th} = \SI{1.21}{\volt}$.
Using Eqs.~\eqref{eq:slowflow}, we can describe the steady-state of the single KPO by applying the condition ($\dot{u}_1=\dot{v}_1=0$)~\cite{Papariello_2016, Leuch_2016,Eichler_Zilberberg_book,kovsata2022harmonicbalance}. This yields a quintic characteristic polynomial with up to three different stable solutions (attractors) in phase space, cf. Fig.~\ref{fig:Fig_Single_Steady}(b). As a function of $f_d$, the number of stable solutions changes at specific bifurcation points. In the single KPO, we only observe pitchfork bifurcations, which involve at least one stable solution, cf. Fig.~\ref{fig:Fig_Single_Steady}(c). 
In Fig.~\ref{fig:Fig_Single_Steady}(d) we show the characteristic parametric instability lobe. Region (i) accommodates only one stable state with amplitude 0.
Inside the region marked as (ii), the linear resonator becomes unstable, bifurcates, and settles into one of the two steady states that are stabilized by $\alpha$~\cite{Lifshitz_Cross}. These phase states have the same amplitude but are $\pi$-shifted in phase, cf. Fig.~\ref{fig:Fig_Single_Steady}(b). In region (iii), the phase states coexist with the amplitude 0 solution.


\textit{Single KPO -- } In Fig.~\ref{fig:Fig_1} we inspect the noise-induced switching of a single KPO, whose properties are well known ~\cite{Dykman_1998, Ryvkine_2006, Chan_2008, Dykman_2018}.
As expected, we find that $\Gamma$ decreases monotonically with increasing separation between the phase states, which we control here through $f_d$~\cite{Dykman_1998}. Similar results have been previously measured in other KPO implementations~\cite{Chan_2008}.

The monotonic decrease of $\Gamma$ in Fig.~\ref{fig:Fig_1}(c) is derived using the Onsager-Machlup approach~\cite{Dykman_1998}. Specifically, at low noise, the switching rate $\Gamma$ is dominated by the path $\mathbf{Y}_{\rm min}$ that minimizes $S_{\rm OM}$~\cite{Lehmann_2003}. 
Repeating this estimation as a function of $f_d$ and calculating $\Gamma_{\rm min}$ yields good agreement with the experimentally observed $\Gamma$, cf. Fig.~\ref{fig:Fig_1}(c). Note that the prefactor $\Gamma_0$ is not obtained by this method but reused from Ref.~\cite{Dykman_1998}, leading to a slight overall shift towards larger $\Gamma$. The analytical formula derived in Ref.~\cite{Dykman_1998} produces a similarly good agreement, cf. Eq.~\eqref{eq:analyticRate} in Appendix~\ref{sec:gamma_det}.

\section{Fluctuating versus Coherent Signal Amplitude}\label{sec:fluctuating_coherent}
To obtain an optimal agreement between the measured switching rates and the theoretical predictions, we consistently found that the noise power spectral density in the model had to be a factor $\approx4.2$ smaller than the value applied in the experiment. This discrepancy is likely due to an additional attenuation of a factor 2 in the path of the fluctuating voltage, for instance a voltage division at a $\SI{50}{\ohm}$ matched input port. The fluctuating signal with power spectral density $S_{n}$ was provided by two dedicated voltage sources with the same output intensity and added to the coherent signal via the ADD channel of the Zurich Instruments HF2LI lock-in amplifier. The resulting noise process $\xi_i$ acting on our system has a power spectral density $\varsigma^2 = C_{in} S_{n}$, where the coefficient for the signal in-coupling efficiency is $C_{in} = \SI{4.93e-20}{\hertz^4}$ for the single-KPO experiment. For the two-KPO experiment, we find best agreement for a slightly lower value for $C_{in}$, which is probably due to differences in the coil geometry between the devices or between the experimental runs.

\begin{figure}[t]
  \includegraphics[width=\linewidth]{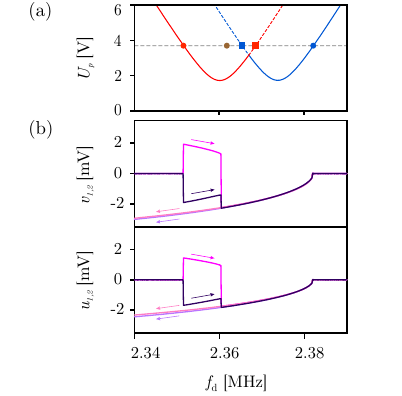}
  \caption{(a)~stability diagram of the system of two coupled KPOs in a space spanned by $f_d$ and $U_p$. The red and blue lines mark the instability lobes of the parametrically driven antisymmetric and symmetric normal modes, respectively. The frequency sweep in Fig.~\ref{fig:Fig_2}(a) and in (b) of this figure was taken at the $U_p$ value indicated by a dashed horizontal line, with dots and squares corresponding to the bifurcations shown in Fig.~\ref{fig:Fig_2}(e). (b)~Full data of the frequency sweep shown in Fig.~\ref{fig:Fig_2}(a). Arrows indicate the sweep direction.} 
  \label{fig:FigS1}
\end{figure}

\begin{figure}[t!]
  \includegraphics[width=\linewidth]{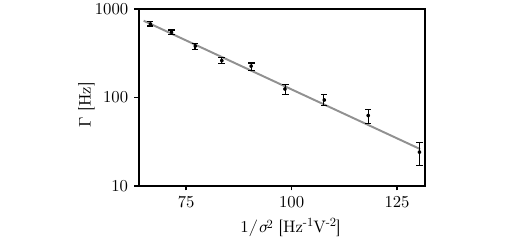}
  \caption{Switching rate $\Gamma$ between the symmetric states of the two-KPO system as a function of inverse noise strength, $1/\sigma^2$, obtained from simulation (black) for $U_d=\SI{3.7}{\volt}$ and $f_d=\SI{2.3725}{\mega\hertz}$. The optimal fit (gray) corresponds to Eq.~\eqref{eq:Gamma} for the fitting parameters $\Gamma_0 = \SI{2e4}{\hertz}$ and $W = \SI{0.024}{\hertz \volt^{2}}$.} 
  \label{fig:FigA5}
\end{figure}

\section{Details on the calculations for noise-induced switching}\label{sec:gamma_det}
\textit{Determination of the switching rate -- } The experimental determination of the switching rate $\Gamma$ in Fig.~\ref{fig:Fig_2}(c) was performed with a lock-in amplifier (Zurich Instruments HF2LI). We used a sampling rate of \SI{450}{\hertz} and a total measurement time of \SI{300}{\second} for $f_d \leq \SI{2.3696}{\hertz}$ and \SI{60}{\second} for $f_d > \SI{2.3696}{\hertz}$. Counting of the switching events was done with a numerical algorithm that compared the amplitudes and phases of successive measurement points for an entire time trace measurement. Concretely, the program increased the switching counter by 1 if the phase difference of two successive points was above a `phase threshold' (\SI{130}{\degree}) while at least one of the points was above an `amplitude threshold' (\SI{0.5}{\milli\volt}), or if exactly one out of two successive points was above the amplitude threshold. The same algorithm was used to evaluate the switching rate in numerical simulations that emulated the measurements (including the effective sampling rate). Similar results were obtained by finding the maximal turning point of Allan deviations of the phase~\cite{margiani2022extracting}.

\begin{figure*}[t!]
  \includegraphics[width=\linewidth]{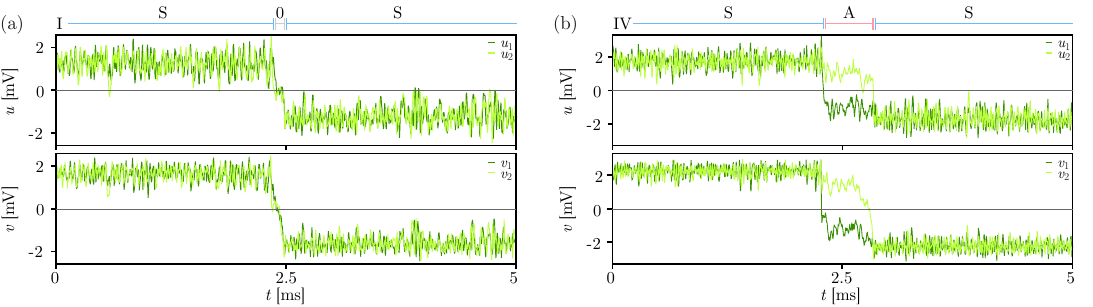}
  \caption{Examples of switching events in the two-KPO system. (a)~For $U_d = \SI{3.7}{\volt}$ and $f_d = \SI{2.37}{\mega\hertz}$ (marked as I in Fig.~\ref{fig:Fig_2}(c)), all four measured coordinates ($u_1, v_1, u_2, v_2$) switch simultaneously on our sampling timescale. The system switches from one symmetric (S) configuration to the opposite one via the 0-amplitude (0) state. (b)~For $U_d = \SI{3.7}{\volt}$ and $f_d = \SI{2.36}{\mega\hertz}$ (marked as IV in Fig.~\ref{fig:Fig_2}(c)), the coordinates ($u_1, v_1$) switch first, followed by ($u_2, v_2$) after a delay of roughly \SI{0.5}{\milli\second}. In the time span between the jumps, the system occupies the antisymmetric (A) state.}
  \label{fig:FigA4}
\end{figure*}

\textit{Determination of the activation barrier $W$ -- }
We  simulate the coupled KPO system at different noise strengths, see Fig~\ref{fig:FigA5}, and we verify that our analysis is in the low-noise limit by showing that Eq.~\eqref{eq:Gamma} is obeyed.
The optimal fit (gray) yields $\Gamma_0 = \SI{2e4}{\hertz}$ and $W = \SI{0.024}{\hertz \volt^{2}}$.
In this limit, $\Gamma_0$ and $W$ are independent of the noise and purely depend on the properties of the non-stochastic system, and on the switching paths in phase space. This procedure allows us to extract $\Gamma_0$ and $W$ for the experimental and the numerical data  in Fig.~\ref{fig:Fig_2}(d), as well as calculate $\Gamma$ at different noise strengths.
For the numerical switching rate in Fig.~\ref{fig:Fig_2}(c), we used this scaling law to convert the numerical data, simulated at $1.3$ times stronger noise, to the experimentally applied noise.

\textit{Analytic expression for a single KPO -- }
For a single parametric Kerr oscillator, the switching rate was calculated in~\cite{Dykman_1998} and is given by
\begin{widetext}
\begin{equation}
\label{eq:analyticRate}
    \Gamma = \frac{ \left(\gamma  \sqrt{\frac{\lambda ^2 \omega_0^4}{\gamma ^2 \omega ^2}-4}-4 \omega +4 \omega_0\right) \sqrt{\left| 1-\frac{\lambda ^2 \omega_0^4}{4 \gamma ^2 \omega ^2}\right| } \exp \left(-\frac{ \gamma ^2 \omega ^3 \left(\gamma  \sqrt{\frac{\lambda ^2 \omega_0^4}{\gamma ^2 \omega ^2}-4}-4 \omega +4 \omega_0\right)^2 \sqrt{\left| 1-\frac{\lambda ^2 \omega_0^4}{4 \gamma ^2 \omega ^2}\right| }}{3 \alpha  \lambda ^2 \sigma ^2 \omega_0^4}\right)}{2 \sqrt{2} \pi   }\,.
\end{equation}
\end{widetext}

\textit{Details on the path optimization -- }
The switching rate $\Gamma$ can be described by $S_{\rm OM}[\mathbf{Y}_{\rm min}]$.  Obtaining the minimal path $\mathbf{Y}_{\rm min}$ is a complex task for coupled parametric resonators. As a simple variation of a discretized path with equal timesteps fails to obtain correct results, we use the sgMAM method~\cite{Grafke_2017}. It is an improved path optimization scheme based on  scaled time, leading to consistent converged results. We start with a guessed initial path that connects two stable states via an unstable attractor. Then, we perform numerical minimization of $S_{\rm OM}$ by varying the path in phase space between the chosen end points.  
For the single KPO, we choose the two phase states as initial and final state, see Fig~\ref{fig:Fig_1}(b). 
In the coupled system with $\mathbf{Y}=(u_1,v_1,u_2,v_2)^T$, we choose one of the symmetric states as the initial point and the other one as the final point, and try different unstable states as intermediate points. We thus obtain the corresponding locally-minimizing switching paths $\mathbf{Y}_{\rm min}$, see Fig.~\ref{fig:Fig_2}(d).

\section{Switching via the antisymmetric state}\label{sec:antisymm_switching}

In Fig.~\ref{fig:FigA4}, we show examples of timetraces during noise-induced switching between symmetric states. Figure~\ref{fig:FigA4}(a) corresponds to $f_d=\SI{2.37}{\mega\hertz}$ in region I of Fig.~\ref{fig:Fig_2}(c), where switches occur via the unstable 0-amplitude state because both resonators switch synchronously. In Fig.~\ref{fig:FigA4}(b), we show an example for $f_d=\SI{2.36}{\mega\hertz}$ in region IV, where the two resonators switch with a finite delay. In the short time interval between the two switches, the system dwells in the antisymmetric state.

\bibliographystyle{apsrev4-1}

\begin{thebibliography}{70}%
\makeatletter
\providecommand \@ifxundefined [1]{%
 \@ifx{#1\undefined}
}%
\providecommand \@ifnum [1]{%
 \ifnum #1\expandafter \@firstoftwo
 \else \expandafter \@secondoftwo
 \fi
}%
\providecommand \@ifx [1]{%
 \ifx #1\expandafter \@firstoftwo
 \else \expandafter \@secondoftwo
 \fi
}%
\providecommand \natexlab [1]{#1}%
\providecommand \enquote  [1]{``#1''}%
\providecommand \bibnamefont  [1]{#1}%
\providecommand \bibfnamefont [1]{#1}%
\providecommand \citenamefont [1]{#1}%
\providecommand \href@noop [0]{\@secondoftwo}%
\providecommand \href [0]{\begingroup \@sanitize@url \@href}%
\providecommand \@href[1]{\@@startlink{#1}\@@href}%
\providecommand \@@href[1]{\endgroup#1\@@endlink}%
\providecommand \@sanitize@url [0]{\catcode `\\12\catcode `\$12\catcode
  `\&12\catcode `\#12\catcode `\^12\catcode `\_12\catcode `\%12\relax}%
\providecommand \@@startlink[1]{}%
\providecommand \@@endlink[0]{}%
\providecommand \url  [0]{\begingroup\@sanitize@url \@url }%
\providecommand \@url [1]{\endgroup\@href {#1}{\urlprefix }}%
\providecommand \urlprefix  [0]{URL }%
\providecommand \Eprint [0]{\href }%
\providecommand \doibase [0]{http://dx.doi.org/}%
\providecommand \selectlanguage [0]{\@gobble}%
\providecommand \bibinfo  [0]{\@secondoftwo}%
\providecommand \bibfield  [0]{\@secondoftwo}%
\providecommand \translation [1]{[#1]}%
\providecommand \BibitemOpen [0]{}%
\providecommand \bibitemStop [0]{}%
\providecommand \bibitemNoStop [0]{.\EOS\space}%
\providecommand \EOS [0]{\spacefactor3000\relax}%
\providecommand \BibitemShut  [1]{\csname bibitem#1\endcsname}%
\let\auto@bib@innerbib\@empty
\bibitem [{\citenamefont {Kramers}(1940)}]{Kramers1940}%
  \BibitemOpen
  \bibfield  {author} {\bibinfo {author} {\bibfnamefont {H.}~\bibnamefont
  {Kramers}},\ }\href {\doibase https://doi.org/10.1016/S0031-8914(40)90098-2}
  {\bibfield  {journal} {\bibinfo  {journal} {Physica}\ }\textbf {\bibinfo
  {volume} {7}},\ \bibinfo {pages} {284} (\bibinfo {year} {1940})}\BibitemShut
  {NoStop}%
\bibitem [{\citenamefont {H\"anggi}\ \emph {et~al.}(1990)\citenamefont
  {H\"anggi}, \citenamefont {Talkner},\ and\ \citenamefont
  {Borkovec}}]{Hanggi_1990}%
  \BibitemOpen
  \bibfield  {author} {\bibinfo {author} {\bibfnamefont {P.}~\bibnamefont
  {H\"anggi}}, \bibinfo {author} {\bibfnamefont {P.}~\bibnamefont {Talkner}}, \
  and\ \bibinfo {author} {\bibfnamefont {M.}~\bibnamefont {Borkovec}},\ }\href
  {\doibase 10.1103/RevModPhys.62.251} {\bibfield  {journal} {\bibinfo
  {journal} {Rev. Mod. Phys.}\ }\textbf {\bibinfo {volume} {62}},\ \bibinfo
  {pages} {251} (\bibinfo {year} {1990})}\BibitemShut {NoStop}%
\bibitem [{\citenamefont {Rondin}\ \emph
  {et~al.}(2017{\natexlab{a}})\citenamefont {Rondin}, \citenamefont {Gieseler},
  \citenamefont {Ricci}, \citenamefont {Quidant}, \citenamefont {Dellago},\
  and\ \citenamefont {Novotny}}]{rondin2017direct}%
  \BibitemOpen
  \bibfield  {author} {\bibinfo {author} {\bibfnamefont {L.}~\bibnamefont
  {Rondin}}, \bibinfo {author} {\bibfnamefont {J.}~\bibnamefont {Gieseler}},
  \bibinfo {author} {\bibfnamefont {F.}~\bibnamefont {Ricci}}, \bibinfo
  {author} {\bibfnamefont {R.}~\bibnamefont {Quidant}}, \bibinfo {author}
  {\bibfnamefont {C.}~\bibnamefont {Dellago}}, \ and\ \bibinfo {author}
  {\bibfnamefont {L.}~\bibnamefont {Novotny}},\ }\href@noop {} {\bibfield
  {journal} {\bibinfo  {journal} {Nature nanotechnology}\ }\textbf {\bibinfo
  {volume} {12}},\ \bibinfo {pages} {1130} (\bibinfo {year}
  {2017}{\natexlab{a}})}\BibitemShut {NoStop}%
\bibitem [{\citenamefont {Best}\ and\ \citenamefont {Hummer}(2006)}]{Best2006}%
  \BibitemOpen
  \bibfield  {author} {\bibinfo {author} {\bibfnamefont {R.~B.}\ \bibnamefont
  {Best}}\ and\ \bibinfo {author} {\bibfnamefont {G.}~\bibnamefont {Hummer}},\
  }\href {\doibase 10.1103/PhysRevLett.96.228104} {\bibfield  {journal}
  {\bibinfo  {journal} {Phys. Rev. Lett.}\ }\textbf {\bibinfo {volume} {96}},\
  \bibinfo {pages} {228104} (\bibinfo {year} {2006})}\BibitemShut {NoStop}%
\bibitem [{\citenamefont {Chung}\ \emph {et~al.}(2015)\citenamefont {Chung},
  \citenamefont {Piana-Agostinetti}, \citenamefont {Shaw},\ and\ \citenamefont
  {Eaton}}]{Chung2015}%
  \BibitemOpen
  \bibfield  {author} {\bibinfo {author} {\bibfnamefont {H.~S.}\ \bibnamefont
  {Chung}}, \bibinfo {author} {\bibfnamefont {S.}~\bibnamefont
  {Piana-Agostinetti}}, \bibinfo {author} {\bibfnamefont {D.~E.}\ \bibnamefont
  {Shaw}}, \ and\ \bibinfo {author} {\bibfnamefont {W.~A.}\ \bibnamefont
  {Eaton}},\ }\href {\doibase 10.1126/science.aab1369} {\bibfield  {journal}
  {\bibinfo  {journal} {Science}\ }\textbf {\bibinfo {volume} {349}},\ \bibinfo
  {pages} {1504} (\bibinfo {year} {2015})},\ \Eprint
  {http://arxiv.org/abs/https://www.science.org/doi/pdf/10.1126/science.aab1369}
  {https://www.science.org/doi/pdf/10.1126/science.aab1369} \BibitemShut
  {NoStop}%
\bibitem [{\citenamefont {Garc\'{\i}a-M\"uller}\ \emph
  {et~al.}(2008)\citenamefont {Garc\'{\i}a-M\"uller}, \citenamefont {Borondo},
  \citenamefont {Hernandez},\ and\ \citenamefont {Benito}}]{GarciaMuller2008}%
  \BibitemOpen
  \bibfield  {author} {\bibinfo {author} {\bibfnamefont {P.~L.}\ \bibnamefont
  {Garc\'{\i}a-M\"uller}}, \bibinfo {author} {\bibfnamefont {F.}~\bibnamefont
  {Borondo}}, \bibinfo {author} {\bibfnamefont {R.}~\bibnamefont {Hernandez}},
  \ and\ \bibinfo {author} {\bibfnamefont {R.~M.}\ \bibnamefont {Benito}},\
  }\href {\doibase 10.1103/PhysRevLett.101.178302} {\bibfield  {journal}
  {\bibinfo  {journal} {Phys. Rev. Lett.}\ }\textbf {\bibinfo {volume} {101}},\
  \bibinfo {pages} {178302} (\bibinfo {year} {2008})}\BibitemShut {NoStop}%
\bibitem [{\citenamefont {Badzey}\ and\ \citenamefont
  {Mohanty}(2005)}]{Badzey2005}%
  \BibitemOpen
  \bibfield  {author} {\bibinfo {author} {\bibfnamefont {R.~L.}\ \bibnamefont
  {Badzey}}\ and\ \bibinfo {author} {\bibfnamefont {P.}~\bibnamefont
  {Mohanty}},\ }\href {\doibase 10.1038/nature04124} {\bibfield  {journal}
  {\bibinfo  {journal} {Nature}\ }\textbf {\bibinfo {volume} {437}},\ \bibinfo
  {pages} {995} (\bibinfo {year} {2005})}\BibitemShut {NoStop}%
\bibitem [{\citenamefont {Rondin}\ \emph
  {et~al.}(2017{\natexlab{b}})\citenamefont {Rondin}, \citenamefont {Gieseler},
  \citenamefont {Ricci}, \citenamefont {Quidant}, \citenamefont {Dellago},\
  and\ \citenamefont {Novotny}}]{Rondin2017}%
  \BibitemOpen
  \bibfield  {author} {\bibinfo {author} {\bibfnamefont {L.}~\bibnamefont
  {Rondin}}, \bibinfo {author} {\bibfnamefont {J.}~\bibnamefont {Gieseler}},
  \bibinfo {author} {\bibfnamefont {F.}~\bibnamefont {Ricci}}, \bibinfo
  {author} {\bibfnamefont {R.}~\bibnamefont {Quidant}}, \bibinfo {author}
  {\bibfnamefont {C.}~\bibnamefont {Dellago}}, \ and\ \bibinfo {author}
  {\bibfnamefont {L.}~\bibnamefont {Novotny}},\ }\href {\doibase
  10.1038/nnano.2017.198} {\bibfield  {journal} {\bibinfo  {journal} {Nature
  Nanotechnology}\ }\textbf {\bibinfo {volume} {12}},\ \bibinfo {pages} {1130}
  (\bibinfo {year} {2017}{\natexlab{b}})}\BibitemShut {NoStop}%
\bibitem [{\citenamefont {Fulton}\ and\ \citenamefont
  {Dunkleberger}(1974)}]{Fulton1974}%
  \BibitemOpen
  \bibfield  {author} {\bibinfo {author} {\bibfnamefont {T.~A.}\ \bibnamefont
  {Fulton}}\ and\ \bibinfo {author} {\bibfnamefont {L.~N.}\ \bibnamefont
  {Dunkleberger}},\ }\href {\doibase 10.1103/PhysRevB.9.4760} {\bibfield
  {journal} {\bibinfo  {journal} {Phys. Rev. B}\ }\textbf {\bibinfo {volume}
  {9}},\ \bibinfo {pages} {4760} (\bibinfo {year} {1974})}\BibitemShut
  {NoStop}%
\bibitem [{\citenamefont {Silvestrini}\ \emph {et~al.}(1988)\citenamefont
  {Silvestrini}, \citenamefont {Pagano}, \citenamefont {Cristiano},
  \citenamefont {Liengme},\ and\ \citenamefont {Gray}}]{Silvestrini1988}%
  \BibitemOpen
  \bibfield  {author} {\bibinfo {author} {\bibfnamefont {P.}~\bibnamefont
  {Silvestrini}}, \bibinfo {author} {\bibfnamefont {S.}~\bibnamefont {Pagano}},
  \bibinfo {author} {\bibfnamefont {R.}~\bibnamefont {Cristiano}}, \bibinfo
  {author} {\bibfnamefont {O.}~\bibnamefont {Liengme}}, \ and\ \bibinfo
  {author} {\bibfnamefont {K.~E.}\ \bibnamefont {Gray}},\ }\href {\doibase
  10.1103/PhysRevLett.60.844} {\bibfield  {journal} {\bibinfo  {journal} {Phys.
  Rev. Lett.}\ }\textbf {\bibinfo {volume} {60}},\ \bibinfo {pages} {844}
  (\bibinfo {year} {1988})}\BibitemShut {NoStop}%
\bibitem [{\citenamefont {Dykman}\ \emph {et~al.}(1998)\citenamefont {Dykman},
  \citenamefont {Maloney}, \citenamefont {Smelyanskiy},\ and\ \citenamefont
  {Silverstein}}]{Dykman_1998}%
  \BibitemOpen
  \bibfield  {author} {\bibinfo {author} {\bibfnamefont {M.~I.}\ \bibnamefont
  {Dykman}}, \bibinfo {author} {\bibfnamefont {C.~M.}\ \bibnamefont {Maloney}},
  \bibinfo {author} {\bibfnamefont {V.~N.}\ \bibnamefont {Smelyanskiy}}, \ and\
  \bibinfo {author} {\bibfnamefont {M.}~\bibnamefont {Silverstein}},\ }\href
  {\doibase 10.1103/PhysRevE.57.5202} {\bibfield  {journal} {\bibinfo
  {journal} {Phys. Rev. E}\ }\textbf {\bibinfo {volume} {57}},\ \bibinfo
  {pages} {5202} (\bibinfo {year} {1998})}\BibitemShut {NoStop}%
\bibitem [{\citenamefont {Dykman}(2007)}]{Dykman_2007}%
  \BibitemOpen
  \bibfield  {author} {\bibinfo {author} {\bibfnamefont {M.~I.}\ \bibnamefont
  {Dykman}},\ }\href {\doibase 10.1103/PhysRevE.75.011101} {\bibfield
  {journal} {\bibinfo  {journal} {Phys. Rev. E}\ }\textbf {\bibinfo {volume}
  {75}},\ \bibinfo {pages} {011101} (\bibinfo {year} {2007})}\BibitemShut
  {NoStop}%
\bibitem [{\citenamefont {Tadokoro}\ \emph {et~al.}(2020)\citenamefont
  {Tadokoro}, \citenamefont {Tanaka},\ and\ \citenamefont
  {Dykman}}]{tadokoro2020noise}%
  \BibitemOpen
  \bibfield  {author} {\bibinfo {author} {\bibfnamefont {Y.}~\bibnamefont
  {Tadokoro}}, \bibinfo {author} {\bibfnamefont {H.}~\bibnamefont {Tanaka}}, \
  and\ \bibinfo {author} {\bibfnamefont {M.}~\bibnamefont {Dykman}},\
  }\href@noop {} {\bibfield  {journal} {\bibinfo  {journal} {Scientific
  Reports}\ }\textbf {\bibinfo {volume} {10}},\ \bibinfo {pages} {10413}
  (\bibinfo {year} {2020})}\BibitemShut {NoStop}%
\bibitem [{\citenamefont {Luchinsky}\ \emph
  {et~al.}(1999{\natexlab{a}})\citenamefont {Luchinsky}, \citenamefont {Maier},
  \citenamefont {Mannella}, \citenamefont {McClintock},\ and\ \citenamefont
  {Stein}}]{Luchinsky1999}%
  \BibitemOpen
  \bibfield  {author} {\bibinfo {author} {\bibfnamefont {D.~G.}\ \bibnamefont
  {Luchinsky}}, \bibinfo {author} {\bibfnamefont {R.~S.}\ \bibnamefont
  {Maier}}, \bibinfo {author} {\bibfnamefont {R.}~\bibnamefont {Mannella}},
  \bibinfo {author} {\bibfnamefont {P.~V.~E.}\ \bibnamefont {McClintock}}, \
  and\ \bibinfo {author} {\bibfnamefont {D.~L.}\ \bibnamefont {Stein}},\ }\href
  {\doibase 10.1103/PhysRevLett.82.1806} {\bibfield  {journal} {\bibinfo
  {journal} {Phys. Rev. Lett.}\ }\textbf {\bibinfo {volume} {82}},\ \bibinfo
  {pages} {1806} (\bibinfo {year} {1999}{\natexlab{a}})}\BibitemShut {NoStop}%
\bibitem [{\citenamefont {Lapidus}\ \emph {et~al.}(1999)\citenamefont
  {Lapidus}, \citenamefont {Enzer},\ and\ \citenamefont {Gabrielse}}]{lapidus}%
  \BibitemOpen
  \bibfield  {author} {\bibinfo {author} {\bibfnamefont {L.~J.}\ \bibnamefont
  {Lapidus}}, \bibinfo {author} {\bibfnamefont {D.}~\bibnamefont {Enzer}}, \
  and\ \bibinfo {author} {\bibfnamefont {G.}~\bibnamefont {Gabrielse}},\ }\href
  {\doibase 10.1103/PhysRevLett.83.899} {\bibfield  {journal} {\bibinfo
  {journal} {Phys. Rev. Lett.}\ }\textbf {\bibinfo {volume} {83}},\ \bibinfo
  {pages} {899} (\bibinfo {year} {1999})}\BibitemShut {NoStop}%
\bibitem [{\citenamefont {Kim}\ \emph {et~al.}(2005)\citenamefont {Kim},
  \citenamefont {Heo}, \citenamefont {Lee}, \citenamefont {Ha}, \citenamefont
  {Jang}, \citenamefont {Noh},\ and\ \citenamefont {Jhe}}]{kkim2005}%
  \BibitemOpen
  \bibfield  {author} {\bibinfo {author} {\bibfnamefont {K.}~\bibnamefont
  {Kim}}, \bibinfo {author} {\bibfnamefont {M.-S.}\ \bibnamefont {Heo}},
  \bibinfo {author} {\bibfnamefont {K.-H.}\ \bibnamefont {Lee}}, \bibinfo
  {author} {\bibfnamefont {H.-J.}\ \bibnamefont {Ha}}, \bibinfo {author}
  {\bibfnamefont {K.}~\bibnamefont {Jang}}, \bibinfo {author} {\bibfnamefont
  {H.-R.}\ \bibnamefont {Noh}}, \ and\ \bibinfo {author} {\bibfnamefont
  {W.}~\bibnamefont {Jhe}},\ }\href {\doibase 10.1103/PhysRevA.72.053402}
  {\bibfield  {journal} {\bibinfo  {journal} {Phys. Rev. A}\ }\textbf {\bibinfo
  {volume} {72}},\ \bibinfo {pages} {053402} (\bibinfo {year}
  {2005})}\BibitemShut {NoStop}%
\bibitem [{\citenamefont {Aldridge}\ and\ \citenamefont
  {Cleland}(2005)}]{Aldridge2005}%
  \BibitemOpen
  \bibfield  {author} {\bibinfo {author} {\bibfnamefont {J.~S.}\ \bibnamefont
  {Aldridge}}\ and\ \bibinfo {author} {\bibfnamefont {A.~N.}\ \bibnamefont
  {Cleland}},\ }\href {\doibase 10.1103/PhysRevLett.94.156403} {\bibfield
  {journal} {\bibinfo  {journal} {Phys. Rev. Lett.}\ }\textbf {\bibinfo
  {volume} {94}},\ \bibinfo {pages} {156403} (\bibinfo {year}
  {2005})}\BibitemShut {NoStop}%
\bibitem [{\citenamefont {Chan}\ and\ \citenamefont
  {Stambaugh}(2007)}]{Chan_2007}%
  \BibitemOpen
  \bibfield  {author} {\bibinfo {author} {\bibfnamefont {H.~B.}\ \bibnamefont
  {Chan}}\ and\ \bibinfo {author} {\bibfnamefont {C.}~\bibnamefont
  {Stambaugh}},\ }\href {\doibase 10.1103/PhysRevLett.99.060601} {\bibfield
  {journal} {\bibinfo  {journal} {Phys. Rev. Lett.}\ }\textbf {\bibinfo
  {volume} {99}},\ \bibinfo {pages} {060601} (\bibinfo {year}
  {2007})}\BibitemShut {NoStop}%
\bibitem [{\citenamefont {Chan}\ \emph {et~al.}(2008)\citenamefont {Chan},
  \citenamefont {Dykman},\ and\ \citenamefont {Stambaugh}}]{Chan_2008}%
  \BibitemOpen
  \bibfield  {author} {\bibinfo {author} {\bibfnamefont {H.~B.}\ \bibnamefont
  {Chan}}, \bibinfo {author} {\bibfnamefont {M.~I.}\ \bibnamefont {Dykman}}, \
  and\ \bibinfo {author} {\bibfnamefont {C.}~\bibnamefont {Stambaugh}},\ }\href
  {\doibase 10.1103/PhysRevLett.100.130602} {\bibfield  {journal} {\bibinfo
  {journal} {Phys. Rev. Lett.}\ }\textbf {\bibinfo {volume} {100}},\ \bibinfo
  {pages} {130602} (\bibinfo {year} {2008})}\BibitemShut {NoStop}%
\bibitem [{\citenamefont {Venstra}\ \emph {et~al.}(2013)\citenamefont
  {Venstra}, \citenamefont {Westra},\ and\ \citenamefont {Van
  Der~Zant}}]{Venstra2013}%
  \BibitemOpen
  \bibfield  {author} {\bibinfo {author} {\bibfnamefont {W.~J.}\ \bibnamefont
  {Venstra}}, \bibinfo {author} {\bibfnamefont {H.~J.}\ \bibnamefont {Westra}},
  \ and\ \bibinfo {author} {\bibfnamefont {H.~S.}\ \bibnamefont {Van
  Der~Zant}},\ }\href@noop {} {\bibfield  {journal} {\bibinfo  {journal}
  {Nature communications}\ }\textbf {\bibinfo {volume} {4}},\ \bibinfo {pages}
  {1} (\bibinfo {year} {2013})}\BibitemShut {NoStop}%
\bibitem [{\citenamefont {Mahboob}\ \emph {et~al.}(2014)\citenamefont
  {Mahboob}, \citenamefont {Mounaix}, \citenamefont {Nishiguchi}, \citenamefont
  {Fujiwara},\ and\ \citenamefont {Yamaguchi}}]{Mahboob_2014_2}%
  \BibitemOpen
  \bibfield  {author} {\bibinfo {author} {\bibfnamefont {I.}~\bibnamefont
  {Mahboob}}, \bibinfo {author} {\bibfnamefont {M.}~\bibnamefont {Mounaix}},
  \bibinfo {author} {\bibfnamefont {K.}~\bibnamefont {Nishiguchi}}, \bibinfo
  {author} {\bibfnamefont {A.}~\bibnamefont {Fujiwara}}, \ and\ \bibinfo
  {author} {\bibfnamefont {H.}~\bibnamefont {Yamaguchi}},\ }\href@noop {}
  {\bibfield  {journal} {\bibinfo  {journal} {Scientific Reports}\ }\textbf
  {\bibinfo {volume} {4}},\ \bibinfo {pages} {4448} (\bibinfo {year}
  {2014})}\BibitemShut {NoStop}%
\bibitem [{\citenamefont {Margiani}\ \emph {et~al.}(2022)\citenamefont
  {Margiani}, \citenamefont {Guerrero}, \citenamefont {Heugel}, \citenamefont
  {Marty}, \citenamefont {Pachlatko}, \citenamefont {Gisler}, \citenamefont
  {Vukasin}, \citenamefont {Kwon}, \citenamefont {Miller}, \citenamefont
  {Bousse} \emph {et~al.}}]{margiani2022extracting}%
  \BibitemOpen
  \bibfield  {author} {\bibinfo {author} {\bibfnamefont {G.}~\bibnamefont
  {Margiani}}, \bibinfo {author} {\bibfnamefont {S.}~\bibnamefont {Guerrero}},
  \bibinfo {author} {\bibfnamefont {T.~L.}\ \bibnamefont {Heugel}}, \bibinfo
  {author} {\bibfnamefont {C.}~\bibnamefont {Marty}}, \bibinfo {author}
  {\bibfnamefont {R.}~\bibnamefont {Pachlatko}}, \bibinfo {author}
  {\bibfnamefont {T.}~\bibnamefont {Gisler}}, \bibinfo {author} {\bibfnamefont
  {G.~D.}\ \bibnamefont {Vukasin}}, \bibinfo {author} {\bibfnamefont {H.-K.}\
  \bibnamefont {Kwon}}, \bibinfo {author} {\bibfnamefont {J.~M.}\ \bibnamefont
  {Miller}}, \bibinfo {author} {\bibfnamefont {N.~E.}\ \bibnamefont {Bousse}},
  \emph {et~al.},\ }\href@noop {} {\bibfield  {journal} {\bibinfo  {journal}
  {Applied Physics Letters}\ }\textbf {\bibinfo {volume} {121}},\ \bibinfo
  {pages} {164101} (\bibinfo {year} {2022})}\BibitemShut {NoStop}%
\bibitem [{\citenamefont {Frimmer}\ \emph {et~al.}(2019)\citenamefont
  {Frimmer}, \citenamefont {Heugel}, \citenamefont {Nosan}, \citenamefont
  {Tebbenjohanns}, \citenamefont {H\"alg}, \citenamefont {Akin}, \citenamefont
  {Degen}, \citenamefont {Novotny}, \citenamefont {Chitra}, \citenamefont
  {Zilberberg},\ and\ \citenamefont {Eichler}}]{Frimmer_2019}%
  \BibitemOpen
  \bibfield  {author} {\bibinfo {author} {\bibfnamefont {M.}~\bibnamefont
  {Frimmer}}, \bibinfo {author} {\bibfnamefont {T.~L.}\ \bibnamefont {Heugel}},
  \bibinfo {author} {\bibfnamefont {i.~c.~v.}\ \bibnamefont {Nosan}}, \bibinfo
  {author} {\bibfnamefont {F.}~\bibnamefont {Tebbenjohanns}}, \bibinfo {author}
  {\bibfnamefont {D.}~\bibnamefont {H\"alg}}, \bibinfo {author} {\bibfnamefont
  {A.}~\bibnamefont {Akin}}, \bibinfo {author} {\bibfnamefont {C.~L.}\
  \bibnamefont {Degen}}, \bibinfo {author} {\bibfnamefont {L.}~\bibnamefont
  {Novotny}}, \bibinfo {author} {\bibfnamefont {R.}~\bibnamefont {Chitra}},
  \bibinfo {author} {\bibfnamefont {O.}~\bibnamefont {Zilberberg}}, \ and\
  \bibinfo {author} {\bibfnamefont {A.}~\bibnamefont {Eichler}},\ }\href
  {\doibase 10.1103/PhysRevLett.123.254102} {\bibfield  {journal} {\bibinfo
  {journal} {Phys. Rev. Lett.}\ }\textbf {\bibinfo {volume} {123}},\ \bibinfo
  {pages} {254102} (\bibinfo {year} {2019})}\BibitemShut {NoStop}%
\bibitem [{\citenamefont {Ryvkine}\ and\ \citenamefont
  {Dykman}(2006)}]{Ryvkine_2006}%
  \BibitemOpen
  \bibfield  {author} {\bibinfo {author} {\bibfnamefont {D.}~\bibnamefont
  {Ryvkine}}\ and\ \bibinfo {author} {\bibfnamefont {M.~I.}\ \bibnamefont
  {Dykman}},\ }\href@noop {} {\bibfield  {journal} {\bibinfo  {journal}
  {Physical Review E}\ }\textbf {\bibinfo {volume} {74}},\ \bibinfo {pages}
  {061118} (\bibinfo {year} {2006})}\BibitemShut {NoStop}%
\bibitem [{\citenamefont {Mahboob}\ and\ \citenamefont
  {Yamaguchi}(2008)}]{Mahboob_2008}%
  \BibitemOpen
  \bibfield  {author} {\bibinfo {author} {\bibfnamefont {I.}~\bibnamefont
  {Mahboob}}\ and\ \bibinfo {author} {\bibfnamefont {H.}~\bibnamefont
  {Yamaguchi}},\ }\href {https://doi.org/10.1038/nnano.2008.84} {\bibfield
  {journal} {\bibinfo  {journal} {Nature Nanotechnology}\ }\textbf {\bibinfo
  {volume} {3}},\ \bibinfo {pages} {275} (\bibinfo {year} {2008})}\BibitemShut
  {NoStop}%
\bibitem [{\citenamefont {Wilson}\ \emph {et~al.}(2010)\citenamefont {Wilson},
  \citenamefont {Duty}, \citenamefont {Sandberg}, \citenamefont {Persson},
  \citenamefont {Shumeiko},\ and\ \citenamefont {Delsing}}]{Wilson_2010}%
  \BibitemOpen
  \bibfield  {author} {\bibinfo {author} {\bibfnamefont {C.~M.}\ \bibnamefont
  {Wilson}}, \bibinfo {author} {\bibfnamefont {T.}~\bibnamefont {Duty}},
  \bibinfo {author} {\bibfnamefont {M.}~\bibnamefont {Sandberg}}, \bibinfo
  {author} {\bibfnamefont {F.}~\bibnamefont {Persson}}, \bibinfo {author}
  {\bibfnamefont {V.}~\bibnamefont {Shumeiko}}, \ and\ \bibinfo {author}
  {\bibfnamefont {P.}~\bibnamefont {Delsing}},\ }\href {\doibase
  10.1103/PhysRevLett.105.233907} {\bibfield  {journal} {\bibinfo  {journal}
  {Phys. Rev. Lett.}\ }\textbf {\bibinfo {volume} {105}},\ \bibinfo {pages}
  {233907} (\bibinfo {year} {2010})}\BibitemShut {NoStop}%
\bibitem [{\citenamefont {Eichler}\ \emph {et~al.}(2011)\citenamefont
  {Eichler}, \citenamefont {Chaste}, \citenamefont {Moser},\ and\ \citenamefont
  {Bachtold}}]{Eichler_2011_NL}%
  \BibitemOpen
  \bibfield  {author} {\bibinfo {author} {\bibfnamefont {A.}~\bibnamefont
  {Eichler}}, \bibinfo {author} {\bibfnamefont {J.}~\bibnamefont {Chaste}},
  \bibinfo {author} {\bibfnamefont {J.}~\bibnamefont {Moser}}, \ and\ \bibinfo
  {author} {\bibfnamefont {A.}~\bibnamefont {Bachtold}},\ }\href {\doibase
  10.1021/nl200950d} {\bibfield  {journal} {\bibinfo  {journal} {Nano Letters}\
  }\textbf {\bibinfo {volume} {11}},\ \bibinfo {pages} {2699} (\bibinfo {year}
  {2011})},\ \bibinfo {note} {pMID: 21615135}\BibitemShut {NoStop}%
\bibitem [{\citenamefont {Leuch}\ \emph {et~al.}(2016)\citenamefont {Leuch},
  \citenamefont {Papariello}, \citenamefont {Zilberberg}, \citenamefont
  {Degen}, \citenamefont {Chitra},\ and\ \citenamefont {Eichler}}]{Leuch_2016}%
  \BibitemOpen
  \bibfield  {author} {\bibinfo {author} {\bibfnamefont {A.}~\bibnamefont
  {Leuch}}, \bibinfo {author} {\bibfnamefont {L.}~\bibnamefont {Papariello}},
  \bibinfo {author} {\bibfnamefont {O.}~\bibnamefont {Zilberberg}}, \bibinfo
  {author} {\bibfnamefont {C.~L.}\ \bibnamefont {Degen}}, \bibinfo {author}
  {\bibfnamefont {R.}~\bibnamefont {Chitra}}, \ and\ \bibinfo {author}
  {\bibfnamefont {A.}~\bibnamefont {Eichler}},\ }\href {\doibase
  10.1103/PhysRevLett.117.214101} {\bibfield  {journal} {\bibinfo  {journal}
  {Phys. Rev. Lett.}\ }\textbf {\bibinfo {volume} {117}},\ \bibinfo {pages}
  {214101} (\bibinfo {year} {2016})}\BibitemShut {NoStop}%
\bibitem [{\citenamefont {Gieseler}\ \emph {et~al.}(2012)\citenamefont
  {Gieseler}, \citenamefont {Deutsch}, \citenamefont {Quidant},\ and\
  \citenamefont {Novotny}}]{Gieseler_2012}%
  \BibitemOpen
  \bibfield  {author} {\bibinfo {author} {\bibfnamefont {J.}~\bibnamefont
  {Gieseler}}, \bibinfo {author} {\bibfnamefont {B.}~\bibnamefont {Deutsch}},
  \bibinfo {author} {\bibfnamefont {R.}~\bibnamefont {Quidant}}, \ and\
  \bibinfo {author} {\bibfnamefont {L.}~\bibnamefont {Novotny}},\ }\href
  {\doibase 10.1103/PhysRevLett.109.103603} {\bibfield  {journal} {\bibinfo
  {journal} {Phys. Rev. Lett.}\ }\textbf {\bibinfo {volume} {109}},\ \bibinfo
  {pages} {103603} (\bibinfo {year} {2012})}\BibitemShut {NoStop}%
\bibitem [{\citenamefont {Lin}\ \emph {et~al.}(2014)\citenamefont {Lin},
  \citenamefont {Inomata}, \citenamefont {Koshino}, \citenamefont {Oliver},
  \citenamefont {Nakamura}, \citenamefont {Tsai},\ and\ \citenamefont
  {Yamamoto}}]{Lin_2014}%
  \BibitemOpen
  \bibfield  {author} {\bibinfo {author} {\bibfnamefont {Z.}~\bibnamefont
  {Lin}}, \bibinfo {author} {\bibfnamefont {K.}~\bibnamefont {Inomata}},
  \bibinfo {author} {\bibfnamefont {K.}~\bibnamefont {Koshino}}, \bibinfo
  {author} {\bibfnamefont {W.~D.}\ \bibnamefont {Oliver}}, \bibinfo {author}
  {\bibfnamefont {Y.}~\bibnamefont {Nakamura}}, \bibinfo {author}
  {\bibfnamefont {J.~S.}\ \bibnamefont {Tsai}}, \ and\ \bibinfo {author}
  {\bibfnamefont {T.}~\bibnamefont {Yamamoto}},\ }\href
  {https://doi.org/10.1038/ncomms5480} {\bibfield  {journal} {\bibinfo
  {journal} {Nature Communications}\ }\textbf {\bibinfo {volume} {5}},\
  \bibinfo {pages} {4480} (\bibinfo {year} {2014})}\BibitemShut {NoStop}%
\bibitem [{\citenamefont {Puri}\ \emph
  {et~al.}(2017{\natexlab{a}})\citenamefont {Puri}, \citenamefont {Boutin},\
  and\ \citenamefont {Blais}}]{Puri_2017}%
  \BibitemOpen
  \bibfield  {author} {\bibinfo {author} {\bibfnamefont {S.}~\bibnamefont
  {Puri}}, \bibinfo {author} {\bibfnamefont {S.}~\bibnamefont {Boutin}}, \ and\
  \bibinfo {author} {\bibfnamefont {A.}~\bibnamefont {Blais}},\ }\href
  {https://doi.org/10.1038/s41534-017-0019-1} {\bibfield  {journal} {\bibinfo
  {journal} {npj Quantum Information}\ }\textbf {\bibinfo {volume} {3}},\
  \bibinfo {pages} {18} (\bibinfo {year} {2017}{\natexlab{a}})}\BibitemShut
  {NoStop}%
\bibitem [{\citenamefont {Eichler}\ \emph {et~al.}(2018)\citenamefont
  {Eichler}, \citenamefont {Heugel}, \citenamefont {Leuch}, \citenamefont
  {Degen}, \citenamefont {Chitra},\ and\ \citenamefont
  {Zilberberg}}]{Eichler_2018}%
  \BibitemOpen
  \bibfield  {author} {\bibinfo {author} {\bibfnamefont {A.}~\bibnamefont
  {Eichler}}, \bibinfo {author} {\bibfnamefont {T.~L.}\ \bibnamefont {Heugel}},
  \bibinfo {author} {\bibfnamefont {A.}~\bibnamefont {Leuch}}, \bibinfo
  {author} {\bibfnamefont {C.~L.}\ \bibnamefont {Degen}}, \bibinfo {author}
  {\bibfnamefont {R.}~\bibnamefont {Chitra}}, \ and\ \bibinfo {author}
  {\bibfnamefont {O.}~\bibnamefont {Zilberberg}},\ }\href {\doibase
  10.1063/1.5031058} {\bibfield  {journal} {\bibinfo  {journal} {Applied
  Physics Letters}\ }\textbf {\bibinfo {volume} {112}},\ \bibinfo {pages}
  {233105} (\bibinfo {year} {2018})},\ \Eprint
  {http://arxiv.org/abs/https://doi.org/10.1063/1.5031058}
  {https://doi.org/10.1063/1.5031058} \BibitemShut {NoStop}%
\bibitem [{\citenamefont {Nosan}\ \emph {et~al.}(2019)\citenamefont {Nosan},
  \citenamefont {M\"arki}, \citenamefont {Hauff}, \citenamefont {Knaut},\ and\
  \citenamefont {Eichler}}]{Nosan_2019}%
  \BibitemOpen
  \bibfield  {author} {\bibinfo {author} {\bibfnamefont {Z.}~\bibnamefont
  {Nosan}}, \bibinfo {author} {\bibfnamefont {P.}~\bibnamefont {M\"arki}},
  \bibinfo {author} {\bibfnamefont {N.}~\bibnamefont {Hauff}}, \bibinfo
  {author} {\bibfnamefont {C.}~\bibnamefont {Knaut}}, \ and\ \bibinfo {author}
  {\bibfnamefont {A.}~\bibnamefont {Eichler}},\ }\href {\doibase
  10.1103/PhysRevE.99.062205} {\bibfield  {journal} {\bibinfo  {journal} {Phys.
  Rev. E}\ }\textbf {\bibinfo {volume} {99}},\ \bibinfo {pages} {062205}
  (\bibinfo {year} {2019})}\BibitemShut {NoStop}%
\bibitem [{\citenamefont {Grimm}\ \emph {et~al.}(2019)\citenamefont {Grimm},
  \citenamefont {Frattini}, \citenamefont {Puri}, \citenamefont {Mundhada},
  \citenamefont {Touzard}, \citenamefont {Mirrahimi}, \citenamefont {Girvin},
  \citenamefont {Shankar},\ and\ \citenamefont {Devoret}}]{Grimm_2019}%
  \BibitemOpen
  \bibfield  {author} {\bibinfo {author} {\bibfnamefont {A.}~\bibnamefont
  {Grimm}}, \bibinfo {author} {\bibfnamefont {N.~E.}\ \bibnamefont {Frattini}},
  \bibinfo {author} {\bibfnamefont {S.}~\bibnamefont {Puri}}, \bibinfo {author}
  {\bibfnamefont {S.~O.}\ \bibnamefont {Mundhada}}, \bibinfo {author}
  {\bibfnamefont {S.}~\bibnamefont {Touzard}}, \bibinfo {author} {\bibfnamefont
  {M.}~\bibnamefont {Mirrahimi}}, \bibinfo {author} {\bibfnamefont {S.~M.}\
  \bibnamefont {Girvin}}, \bibinfo {author} {\bibfnamefont {S.}~\bibnamefont
  {Shankar}}, \ and\ \bibinfo {author} {\bibfnamefont {M.~H.}\ \bibnamefont
  {Devoret}},\ }\href {\doibase 10.1038/s41586-020-2587-z} {\bibfield
  {journal} {\bibinfo  {journal} {Nature}\ }\textbf {\bibinfo {volume} {584}},\
  \bibinfo {pages} {205} (\bibinfo {year} {2019})}\BibitemShut {NoStop}%
\bibitem [{\citenamefont {Puri}\ \emph {et~al.}(2019)\citenamefont {Puri},
  \citenamefont {Grimm}, \citenamefont {Campagne-Ibarcq}, \citenamefont
  {Eickbusch}, \citenamefont {Noh}, \citenamefont {Roberts}, \citenamefont
  {Jiang}, \citenamefont {Mirrahimi}, \citenamefont {Devoret},\ and\
  \citenamefont {Girvin}}]{Puri_2019_PRX}%
  \BibitemOpen
  \bibfield  {author} {\bibinfo {author} {\bibfnamefont {S.}~\bibnamefont
  {Puri}}, \bibinfo {author} {\bibfnamefont {A.}~\bibnamefont {Grimm}},
  \bibinfo {author} {\bibfnamefont {P.}~\bibnamefont {Campagne-Ibarcq}},
  \bibinfo {author} {\bibfnamefont {A.}~\bibnamefont {Eickbusch}}, \bibinfo
  {author} {\bibfnamefont {K.}~\bibnamefont {Noh}}, \bibinfo {author}
  {\bibfnamefont {G.}~\bibnamefont {Roberts}}, \bibinfo {author} {\bibfnamefont
  {L.}~\bibnamefont {Jiang}}, \bibinfo {author} {\bibfnamefont
  {M.}~\bibnamefont {Mirrahimi}}, \bibinfo {author} {\bibfnamefont {M.~H.}\
  \bibnamefont {Devoret}}, \ and\ \bibinfo {author} {\bibfnamefont {S.~M.}\
  \bibnamefont {Girvin}},\ }\href {\doibase 10.1103/PhysRevX.9.041009}
  {\bibfield  {journal} {\bibinfo  {journal} {Phys. Rev. X}\ }\textbf {\bibinfo
  {volume} {9}},\ \bibinfo {pages} {041009} (\bibinfo {year}
  {2019})}\BibitemShut {NoStop}%
\bibitem [{\citenamefont {Miller}\ \emph {et~al.}(2019)\citenamefont {Miller},
  \citenamefont {Shin}, \citenamefont {Kwon}, \citenamefont {Shaw},\ and\
  \citenamefont {Kenny}}]{Miller_2019_phase}%
  \BibitemOpen
  \bibfield  {author} {\bibinfo {author} {\bibfnamefont {J.~M.}\ \bibnamefont
  {Miller}}, \bibinfo {author} {\bibfnamefont {D.~D.}\ \bibnamefont {Shin}},
  \bibinfo {author} {\bibfnamefont {H.-K.}\ \bibnamefont {Kwon}}, \bibinfo
  {author} {\bibfnamefont {S.~W.}\ \bibnamefont {Shaw}}, \ and\ \bibinfo
  {author} {\bibfnamefont {T.~W.}\ \bibnamefont {Kenny}},\ }\href {\doibase
  10.1103/PhysRevApplied.12.044053} {\bibfield  {journal} {\bibinfo  {journal}
  {Phys. Rev. Applied}\ }\textbf {\bibinfo {volume} {12}},\ \bibinfo {pages}
  {044053} (\bibinfo {year} {2019})}\BibitemShut {NoStop}%
\bibitem [{\citenamefont {Dykman}(2012)}]{DykmanBook}%
  \BibitemOpen
  \bibfield  {author} {\bibinfo {author} {\bibfnamefont {M.}~\bibnamefont
  {Dykman}},\ }\href@noop {} {\emph {\bibinfo {title} {Fluctuating Nonlinear
  Oscillators}}}\ (\bibinfo  {publisher} {Oxford University Press},\ \bibinfo
  {year} {2012})\BibitemShut {NoStop}%
\bibitem [{\citenamefont {Eichler}\ and\ \citenamefont
  {Zilberberg}(2023)}]{Eichler_Zilberberg_book}%
  \BibitemOpen
  \bibfield  {author} {\bibinfo {author} {\bibfnamefont {A.}~\bibnamefont
  {Eichler}}\ and\ \bibinfo {author} {\bibfnamefont {O.}~\bibnamefont
  {Zilberberg}},\ }\href@noop {} {\emph {\bibinfo {title} {Classical and
  Quantum Parametric Phenomena}}}\ (\bibinfo  {publisher} {Oxford University
  Press},\ \bibinfo {year} {2023})\BibitemShut {NoStop}%
\bibitem [{\citenamefont {Mahboob}\ \emph {et~al.}(2016)\citenamefont
  {Mahboob}, \citenamefont {Okamoto},\ and\ \citenamefont
  {Yamaguchi}}]{Mahboob_2016}%
  \BibitemOpen
  \bibfield  {author} {\bibinfo {author} {\bibfnamefont {I.}~\bibnamefont
  {Mahboob}}, \bibinfo {author} {\bibfnamefont {H.}~\bibnamefont {Okamoto}}, \
  and\ \bibinfo {author} {\bibfnamefont {H.}~\bibnamefont {Yamaguchi}},\ }\href
  {https://advances.sciencemag.org/content/2/6/e1600236} {\bibfield  {journal}
  {\bibinfo  {journal} {Science Advances}\ }\textbf {\bibinfo {volume} {2}},\
  \bibinfo {pages} {e1600236} (\bibinfo {year} {2016})}\BibitemShut {NoStop}%
\bibitem [{\citenamefont {Inagaki}\ \emph {et~al.}(2016)\citenamefont
  {Inagaki}, \citenamefont {Inaba}, \citenamefont {Hamerly}, \citenamefont
  {Inoue}, \citenamefont {Yamamoto},\ and\ \citenamefont
  {Takesue}}]{Inagaki_2016}%
  \BibitemOpen
  \bibfield  {author} {\bibinfo {author} {\bibfnamefont {T.}~\bibnamefont
  {Inagaki}}, \bibinfo {author} {\bibfnamefont {K.}~\bibnamefont {Inaba}},
  \bibinfo {author} {\bibfnamefont {R.}~\bibnamefont {Hamerly}}, \bibinfo
  {author} {\bibfnamefont {K.}~\bibnamefont {Inoue}}, \bibinfo {author}
  {\bibfnamefont {Y.}~\bibnamefont {Yamamoto}}, \ and\ \bibinfo {author}
  {\bibfnamefont {H.}~\bibnamefont {Takesue}},\ }\href
  {https://doi.org/10.1038/nphoton.2016.68} {\bibfield  {journal} {\bibinfo
  {journal} {Nature Photonics}\ }\textbf {\bibinfo {volume} {10}},\ \bibinfo
  {pages} {415} (\bibinfo {year} {2016})}\BibitemShut {NoStop}%
\bibitem [{\citenamefont {Goto}(2016)}]{Goto_2016}%
  \BibitemOpen
  \bibfield  {author} {\bibinfo {author} {\bibfnamefont {H.}~\bibnamefont
  {Goto}},\ }\href {https://doi.org/10.1038/srep21686} {\bibfield  {journal}
  {\bibinfo  {journal} {Scientific Reports}\ }\textbf {\bibinfo {volume} {6}},\
  \bibinfo {pages} {21686} (\bibinfo {year} {2016})}\BibitemShut {NoStop}%
\bibitem [{\citenamefont {Puri}\ \emph
  {et~al.}(2017{\natexlab{b}})\citenamefont {Puri}, \citenamefont {Andersen},
  \citenamefont {Grimsmo},\ and\ \citenamefont {Blais}}]{Puri_2017_NC}%
  \BibitemOpen
  \bibfield  {author} {\bibinfo {author} {\bibfnamefont {S.}~\bibnamefont
  {Puri}}, \bibinfo {author} {\bibfnamefont {C.~K.}\ \bibnamefont {Andersen}},
  \bibinfo {author} {\bibfnamefont {A.~L.}\ \bibnamefont {Grimsmo}}, \ and\
  \bibinfo {author} {\bibfnamefont {A.}~\bibnamefont {Blais}},\ }\href
  {\doibase 10.1038/ncomms15785} {\bibfield  {journal} {\bibinfo  {journal}
  {Nature Communications}\ }\textbf {\bibinfo {volume} {8}},\ \bibinfo {pages}
  {15785} (\bibinfo {year} {2017}{\natexlab{b}})}\BibitemShut {NoStop}%
\bibitem [{\citenamefont {Nigg}\ \emph {et~al.}(2017)\citenamefont {Nigg},
  \citenamefont {L{\"o}rch},\ and\ \citenamefont {Tiwari}}]{Nigg_2017}%
  \BibitemOpen
  \bibfield  {author} {\bibinfo {author} {\bibfnamefont {S.~E.}\ \bibnamefont
  {Nigg}}, \bibinfo {author} {\bibfnamefont {N.}~\bibnamefont {L{\"o}rch}}, \
  and\ \bibinfo {author} {\bibfnamefont {R.~P.}\ \bibnamefont {Tiwari}},\
  }\href {\doibase 10.1126/sciadv.1602273} {\bibfield  {journal} {\bibinfo
  {journal} {Science Advances}\ }\textbf {\bibinfo {volume} {3}} (\bibinfo
  {year} {2017}),\ 10.1126/sciadv.1602273}\BibitemShut {NoStop}%
\bibitem [{\citenamefont {Dykman}\ \emph {et~al.}(2018)\citenamefont {Dykman},
  \citenamefont {Bruder}, \citenamefont {L\"orch},\ and\ \citenamefont
  {Zhang}}]{Dykman_2018}%
  \BibitemOpen
  \bibfield  {author} {\bibinfo {author} {\bibfnamefont {M.~I.}\ \bibnamefont
  {Dykman}}, \bibinfo {author} {\bibfnamefont {C.}~\bibnamefont {Bruder}},
  \bibinfo {author} {\bibfnamefont {N.}~\bibnamefont {L\"orch}}, \ and\
  \bibinfo {author} {\bibfnamefont {Y.}~\bibnamefont {Zhang}},\ }\href
  {\doibase 10.1103/PhysRevB.98.195444} {\bibfield  {journal} {\bibinfo
  {journal} {Phys. Rev. B}\ }\textbf {\bibinfo {volume} {98}},\ \bibinfo
  {pages} {195444} (\bibinfo {year} {2018})}\BibitemShut {NoStop}%
\bibitem [{\citenamefont {Okawachi}\ \emph {et~al.}(2020)\citenamefont
  {Okawachi}, \citenamefont {Yu}, \citenamefont {Jang}, \citenamefont {Ji},
  \citenamefont {Zhao}, \citenamefont {Kim}, \citenamefont {Lipson},\ and\
  \citenamefont {Gaeta}}]{Okawachi_2020}%
  \BibitemOpen
  \bibfield  {author} {\bibinfo {author} {\bibfnamefont {Y.}~\bibnamefont
  {Okawachi}}, \bibinfo {author} {\bibfnamefont {M.}~\bibnamefont {Yu}},
  \bibinfo {author} {\bibfnamefont {J.~K.}\ \bibnamefont {Jang}}, \bibinfo
  {author} {\bibfnamefont {X.}~\bibnamefont {Ji}}, \bibinfo {author}
  {\bibfnamefont {Y.}~\bibnamefont {Zhao}}, \bibinfo {author} {\bibfnamefont
  {B.~Y.}\ \bibnamefont {Kim}}, \bibinfo {author} {\bibfnamefont
  {M.}~\bibnamefont {Lipson}}, \ and\ \bibinfo {author} {\bibfnamefont {A.~L.}\
  \bibnamefont {Gaeta}},\ }\href {\doibase 10.1038/s41467-020-17919-6}
  {\bibfield  {journal} {\bibinfo  {journal} {Nature Communications}\ }\textbf
  {\bibinfo {volume} {11}},\ \bibinfo {pages} {4119} (\bibinfo {year}
  {2020})}\BibitemShut {NoStop}%
\bibitem [{\citenamefont {Calvanese~Strinati}\ \emph
  {et~al.}(2021)\citenamefont {Calvanese~Strinati}, \citenamefont {Bello},
  \citenamefont {Dalla~Torre},\ and\ \citenamefont {Pe'er}}]{Strinanti_2021}%
  \BibitemOpen
  \bibfield  {author} {\bibinfo {author} {\bibfnamefont {M.}~\bibnamefont
  {Calvanese~Strinati}}, \bibinfo {author} {\bibfnamefont {L.}~\bibnamefont
  {Bello}}, \bibinfo {author} {\bibfnamefont {E.~G.}\ \bibnamefont
  {Dalla~Torre}}, \ and\ \bibinfo {author} {\bibfnamefont {A.}~\bibnamefont
  {Pe'er}},\ }\href {\doibase 10.1103/PhysRevLett.126.143901} {\bibfield
  {journal} {\bibinfo  {journal} {Phys. Rev. Lett.}\ }\textbf {\bibinfo
  {volume} {126}},\ \bibinfo {pages} {143901} (\bibinfo {year}
  {2021})}\BibitemShut {NoStop}%
\bibitem [{\citenamefont {Bello}\ \emph {et~al.}(2019)\citenamefont {Bello},
  \citenamefont {Calvanese~Strinati}, \citenamefont {Dalla~Torre},\ and\
  \citenamefont {Pe'er}}]{Bello_2019}%
  \BibitemOpen
  \bibfield  {author} {\bibinfo {author} {\bibfnamefont {L.}~\bibnamefont
  {Bello}}, \bibinfo {author} {\bibfnamefont {M.}~\bibnamefont
  {Calvanese~Strinati}}, \bibinfo {author} {\bibfnamefont {E.~G.}\ \bibnamefont
  {Dalla~Torre}}, \ and\ \bibinfo {author} {\bibfnamefont {A.}~\bibnamefont
  {Pe'er}},\ }\href {\doibase 10.1103/PhysRevLett.123.083901} {\bibfield
  {journal} {\bibinfo  {journal} {Phys. Rev. Lett.}\ }\textbf {\bibinfo
  {volume} {123}},\ \bibinfo {pages} {083901} (\bibinfo {year}
  {2019})}\BibitemShut {NoStop}%
\bibitem [{\citenamefont {Heugel}\ \emph
  {et~al.}(2022{\natexlab{a}})\citenamefont {Heugel}, \citenamefont
  {Zilberberg}, \citenamefont {Marty}, \citenamefont {Chitra},\ and\
  \citenamefont {Eichler}}]{Heugel_2022}%
  \BibitemOpen
  \bibfield  {author} {\bibinfo {author} {\bibfnamefont {T.~L.}\ \bibnamefont
  {Heugel}}, \bibinfo {author} {\bibfnamefont {O.}~\bibnamefont {Zilberberg}},
  \bibinfo {author} {\bibfnamefont {C.}~\bibnamefont {Marty}}, \bibinfo
  {author} {\bibfnamefont {R.}~\bibnamefont {Chitra}}, \ and\ \bibinfo {author}
  {\bibfnamefont {A.}~\bibnamefont {Eichler}},\ }\href@noop {} {\bibfield
  {journal} {\bibinfo  {journal} {Physical Review Research}\ }\textbf {\bibinfo
  {volume} {4}},\ \bibinfo {pages} {013149} (\bibinfo {year}
  {2022}{\natexlab{a}})}\BibitemShut {NoStop}%
\bibitem [{\citenamefont {Heugel}\ \emph
  {et~al.}(2022{\natexlab{b}})\citenamefont {Heugel}, \citenamefont {Eichler},
  \citenamefont {Chitra},\ and\ \citenamefont {Zilberberg}}]{heugel2022role}%
  \BibitemOpen
  \bibfield  {author} {\bibinfo {author} {\bibfnamefont {T.~L.}\ \bibnamefont
  {Heugel}}, \bibinfo {author} {\bibfnamefont {A.}~\bibnamefont {Eichler}},
  \bibinfo {author} {\bibfnamefont {R.}~\bibnamefont {Chitra}}, \ and\ \bibinfo
  {author} {\bibfnamefont {O.}~\bibnamefont {Zilberberg}},\ }\href@noop {}
  {\bibfield  {journal} {\bibinfo  {journal} {arXiv preprint arXiv:2203.05577}\
  } (\bibinfo {year} {2022}{\natexlab{b}})}\BibitemShut {NoStop}%
\bibitem [{\citenamefont {Margiani}\ \emph {et~al.}(2023)\citenamefont
  {Margiani}, \citenamefont {del Pino}, \citenamefont {Heugel}, \citenamefont
  {Bousse}, \citenamefont {Guerrero}, \citenamefont {Kenny}, \citenamefont
  {Zilberberg}, \citenamefont {Sabonis},\ and\ \citenamefont
  {Eichler}}]{margiani2022deterministic}%
  \BibitemOpen
  \bibfield  {author} {\bibinfo {author} {\bibfnamefont {G.}~\bibnamefont
  {Margiani}}, \bibinfo {author} {\bibfnamefont {J.}~\bibnamefont {del Pino}},
  \bibinfo {author} {\bibfnamefont {T.~L.}\ \bibnamefont {Heugel}}, \bibinfo
  {author} {\bibfnamefont {N.~E.}\ \bibnamefont {Bousse}}, \bibinfo {author}
  {\bibfnamefont {S.}~\bibnamefont {Guerrero}}, \bibinfo {author}
  {\bibfnamefont {T.~W.}\ \bibnamefont {Kenny}}, \bibinfo {author}
  {\bibfnamefont {O.}~\bibnamefont {Zilberberg}}, \bibinfo {author}
  {\bibfnamefont {D.}~\bibnamefont {Sabonis}}, \ and\ \bibinfo {author}
  {\bibfnamefont {A.}~\bibnamefont {Eichler}},\ }\href {\doibase
  10.1103/PhysRevResearch.5.L012029} {\bibfield  {journal} {\bibinfo  {journal}
  {Phys. Rev. Res.}\ }\textbf {\bibinfo {volume} {5}},\ \bibinfo {pages}
  {L012029} (\bibinfo {year} {2023})}\BibitemShut {NoStop}%
\bibitem [{\citenamefont {Heugel}\ \emph {et~al.}(2019)\citenamefont {Heugel},
  \citenamefont {Oscity}, \citenamefont {Eichler}, \citenamefont {Zilberberg},\
  and\ \citenamefont {Chitra}}]{Heugel_2019_TC}%
  \BibitemOpen
  \bibfield  {author} {\bibinfo {author} {\bibfnamefont {T.~L.}\ \bibnamefont
  {Heugel}}, \bibinfo {author} {\bibfnamefont {M.}~\bibnamefont {Oscity}},
  \bibinfo {author} {\bibfnamefont {A.}~\bibnamefont {Eichler}}, \bibinfo
  {author} {\bibfnamefont {O.}~\bibnamefont {Zilberberg}}, \ and\ \bibinfo
  {author} {\bibfnamefont {R.}~\bibnamefont {Chitra}},\ }\href {\doibase
  10.1103/PhysRevLett.123.124301} {\bibfield  {journal} {\bibinfo  {journal}
  {Phys. Rev. Lett.}\ }\textbf {\bibinfo {volume} {123}},\ \bibinfo {pages}
  {124301} (\bibinfo {year} {2019})}\BibitemShut {NoStop}%
\bibitem [{\citenamefont {Lehmann}\ \emph {et~al.}(2003)\citenamefont
  {Lehmann}, \citenamefont {Reimann},\ and\ \citenamefont
  {Hänggi}}]{Lehmann_2003}%
  \BibitemOpen
  \bibfield  {author} {\bibinfo {author} {\bibfnamefont {J.}~\bibnamefont
  {Lehmann}}, \bibinfo {author} {\bibfnamefont {P.}~\bibnamefont {Reimann}}, \
  and\ \bibinfo {author} {\bibfnamefont {P.}~\bibnamefont {Hänggi}},\ }\href
  {\doibase https://doi.org/10.1002/pssb.200301774} {\bibfield  {journal}
  {\bibinfo  {journal} {physica status solidi (b)}\ }\textbf {\bibinfo {volume}
  {237}},\ \bibinfo {pages} {53} (\bibinfo {year} {2003})},\ \Eprint
  {http://arxiv.org/abs/https://onlinelibrary.wiley.com/doi/pdf/10.1002/pssb.200301774}
  {https://onlinelibrary.wiley.com/doi/pdf/10.1002/pssb.200301774} \BibitemShut
  {NoStop}%
\bibitem [{\citenamefont {Wio}(2013)}]{Wio_2013}%
  \BibitemOpen
  \bibfield  {author} {\bibinfo {author} {\bibfnamefont {H.~S.}\ \bibnamefont
  {Wio}},\ }\href {\doibase 10.1142/8695} {\emph {\bibinfo {title} {Path
  Integrals for Stochastic Processes}}}\ (\bibinfo  {publisher} {WORLD
  SCIENTIFIC},\ \bibinfo {year} {2013})\ \Eprint
  {http://arxiv.org/abs/https://www.worldscientific.com/doi/pdf/10.1142/8695}
  {https://www.worldscientific.com/doi/pdf/10.1142/8695} \BibitemShut {NoStop}%
\bibitem [{\citenamefont {McLachlan}(1951)}]{mclachlan1951theory}%
  \BibitemOpen
  \bibfield  {author} {\bibinfo {author} {\bibfnamefont {N.}~\bibnamefont
  {McLachlan}},\ }\href@noop {} {\emph {\bibinfo {title} {Theory and
  application of Mathieu functions}}}\ (\bibinfo  {publisher} {Clarendon},\
  \bibinfo {year} {1951})\BibitemShut {NoStop}%
\bibitem [{\citenamefont {Lifshitz}(2009)}]{Lifshitz_Cross}%
  \BibitemOpen
  \bibfield  {author} {\bibinfo {author} {\bibfnamefont {M.~C.}\ \bibnamefont
  {Lifshitz}, \bibfnamefont {R.~Cross}},\ }\enquote {\bibinfo {title}
  {Nonlinear dynamics of nanomechanical and micromechanical resonators},}\ in\
  \href {\doibase 10.1002/9783527626359.ch1} {\emph {\bibinfo {booktitle}
  {Reviews of Nonlinear Dynamics and Complexity}}}\ (\bibinfo  {publisher}
  {Wiley-VCH},\ \bibinfo {year} {2009})\ pp.\ \bibinfo {pages}
  {1--52}\BibitemShut {NoStop}%
\bibitem [{\citenamefont {Guckenheimer}\ and\ \citenamefont
  {Holmes}(1990)}]{guckenheimer_1990}%
  \BibitemOpen
  \bibfield  {author} {\bibinfo {author} {\bibfnamefont {J.}~\bibnamefont
  {Guckenheimer}}\ and\ \bibinfo {author} {\bibfnamefont {P.}~\bibnamefont
  {Holmes}},\ }\href@noop {} {\emph {\bibinfo {title} {Nonlinear oscillations,
  dynamical systems, and bifurcations of vector fields}}},\ Applied
  mathematical sciences\ (\bibinfo  {publisher} {Springer-Verlag},\ \bibinfo
  {year} {1990})\BibitemShut {NoStop}%
\bibitem [{\citenamefont {Papariello}\ \emph {et~al.}(2016)\citenamefont
  {Papariello}, \citenamefont {Zilberberg}, \citenamefont {Eichler},\ and\
  \citenamefont {Chitra}}]{Papariello_2016}%
  \BibitemOpen
  \bibfield  {author} {\bibinfo {author} {\bibfnamefont {L.}~\bibnamefont
  {Papariello}}, \bibinfo {author} {\bibfnamefont {O.}~\bibnamefont
  {Zilberberg}}, \bibinfo {author} {\bibfnamefont {A.}~\bibnamefont {Eichler}},
  \ and\ \bibinfo {author} {\bibfnamefont {R.}~\bibnamefont {Chitra}},\
  }\href@noop {} {\bibfield  {journal} {\bibinfo  {journal} {Phys. Rev. E}\
  }\textbf {\bibinfo {volume} {94}},\ \bibinfo {pages} {022201} (\bibinfo
  {year} {2016})}\BibitemShut {NoStop}%
\bibitem [{\citenamefont {Khas’minskii}(1966)}]{Khasminskii_66}%
  \BibitemOpen
  \bibfield  {author} {\bibinfo {author} {\bibfnamefont {R.~Z.}\ \bibnamefont
  {Khas’minskii}},\ }\href {\doibase 10.1137/1111038} {\bibfield  {journal}
  {\bibinfo  {journal} {Theory of Probability \& Its Applications}\ }\textbf
  {\bibinfo {volume} {11}},\ \bibinfo {pages} {390} (\bibinfo {year} {1966})},\
  \Eprint {http://arxiv.org/abs/https://doi.org/10.1137/1111038}
  {https://doi.org/10.1137/1111038} \BibitemShut {NoStop}%
\bibitem [{\citenamefont {Roberts}\ and\ \citenamefont
  {Spanos}(1986)}]{Roberts_86}%
  \BibitemOpen
  \bibfield  {author} {\bibinfo {author} {\bibfnamefont {J.}~\bibnamefont
  {Roberts}}\ and\ \bibinfo {author} {\bibfnamefont {P.}~\bibnamefont
  {Spanos}},\ }\href {\doibase https://doi.org/10.1016/0020-7462(86)90025-9}
  {\bibfield  {journal} {\bibinfo  {journal} {International Journal of
  Non-Linear Mechanics}\ }\textbf {\bibinfo {volume} {21}},\ \bibinfo {pages}
  {111 } (\bibinfo {year} {1986})}\BibitemShut {NoStop}%
\bibitem [{\citenamefont {Nayfeh}\ and\ \citenamefont
  {Mook}(2008)}]{nayfeh2008}%
  \BibitemOpen
  \bibfield  {author} {\bibinfo {author} {\bibfnamefont {A.~H.}\ \bibnamefont
  {Nayfeh}}\ and\ \bibinfo {author} {\bibfnamefont {D.~T.}\ \bibnamefont
  {Mook}},\ }\href@noop {} {\emph {\bibinfo {title} {Nonlinear
  Oscillations}}},\ Physics textbook\ (\bibinfo  {publisher} {Wiley},\ \bibinfo
  {year} {2008})\BibitemShut {NoStop}%
\bibitem [{\citenamefont {Stambaugh}\ and\ \citenamefont
  {Chan}(2006)}]{Stambaugh_2006}%
  \BibitemOpen
  \bibfield  {author} {\bibinfo {author} {\bibfnamefont {C.}~\bibnamefont
  {Stambaugh}}\ and\ \bibinfo {author} {\bibfnamefont {H.~B.}\ \bibnamefont
  {Chan}},\ }\href {\doibase 10.1103/PhysRevB.73.172302} {\bibfield  {journal}
  {\bibinfo  {journal} {Phys. Rev. B}\ }\textbf {\bibinfo {volume} {73}},\
  \bibinfo {pages} {172302} (\bibinfo {year} {2006})}\BibitemShut {NoStop}%
\bibitem [{\citenamefont {Grafke}\ \emph {et~al.}(2017)\citenamefont {Grafke},
  \citenamefont {Sch{\"a}fer},\ and\ \citenamefont
  {Vanden-Eijnden}}]{Grafke_2017}%
  \BibitemOpen
  \bibfield  {author} {\bibinfo {author} {\bibfnamefont {T.}~\bibnamefont
  {Grafke}}, \bibinfo {author} {\bibfnamefont {T.}~\bibnamefont {Sch{\"a}fer}},
  \ and\ \bibinfo {author} {\bibfnamefont {E.}~\bibnamefont {Vanden-Eijnden}},\
  }\enquote {\bibinfo {title} {Long term effects of small random perturbations
  on dynamical systems: Theoretical and computational tools},}\ in\ \href
  {\doibase 10.1007/978-1-4939-6969-2_2} {\emph {\bibinfo {booktitle} {Recent
  Progress and Modern Challenges in Applied Mathematics, Modeling and
  Computational Science}}},\ \bibinfo {editor} {edited by\ \bibinfo {editor}
  {\bibfnamefont {R.}~\bibnamefont {Melnik}}, \bibinfo {editor} {\bibfnamefont
  {R.}~\bibnamefont {Makarov}}, \ and\ \bibinfo {editor} {\bibfnamefont
  {J.}~\bibnamefont {Belair}}}\ (\bibinfo  {publisher} {Springer New York},\
  \bibinfo {address} {New York, NY},\ \bibinfo {year} {2017})\ pp.\ \bibinfo
  {pages} {17--55}\BibitemShut {NoStop}%
\bibitem [{\citenamefont {Maier}\ and\ \citenamefont
  {Stein}(1992)}]{Maier_1992}%
  \BibitemOpen
  \bibfield  {author} {\bibinfo {author} {\bibfnamefont {R.~S.}\ \bibnamefont
  {Maier}}\ and\ \bibinfo {author} {\bibfnamefont {D.~L.}\ \bibnamefont
  {Stein}},\ }\href {\doibase 10.1103/PhysRevLett.69.3691} {\bibfield
  {journal} {\bibinfo  {journal} {Phys. Rev. Lett.}\ }\textbf {\bibinfo
  {volume} {69}},\ \bibinfo {pages} {3691} (\bibinfo {year}
  {1992})}\BibitemShut {NoStop}%
\bibitem [{\citenamefont {Tang}\ \emph {et~al.}(2017)\citenamefont {Tang},
  \citenamefont {Yuan}, \citenamefont {Wang}, \citenamefont {Zhu},\ and\
  \citenamefont {Ao}}]{Tang_2017}%
  \BibitemOpen
  \bibfield  {author} {\bibinfo {author} {\bibfnamefont {Y.}~\bibnamefont
  {Tang}}, \bibinfo {author} {\bibfnamefont {R.}~\bibnamefont {Yuan}}, \bibinfo
  {author} {\bibfnamefont {G.}~\bibnamefont {Wang}}, \bibinfo {author}
  {\bibfnamefont {X.}~\bibnamefont {Zhu}}, \ and\ \bibinfo {author}
  {\bibfnamefont {P.}~\bibnamefont {Ao}},\ }\href {\doibase
  10.1038/s41598-017-15889-2} {\bibfield  {journal} {\bibinfo  {journal}
  {Scientific Reports}\ }\textbf {\bibinfo {volume} {7}},\ \bibinfo {pages}
  {15762} (\bibinfo {year} {2017})}\BibitemShut {NoStop}%
\bibitem [{\citenamefont {Luchinsky}\ \emph
  {et~al.}(1999{\natexlab{b}})\citenamefont {Luchinsky}, \citenamefont {Maier},
  \citenamefont {Mannella}, \citenamefont {McClintock},\ and\ \citenamefont
  {Stein}}]{Luchinsky_1999}%
  \BibitemOpen
  \bibfield  {author} {\bibinfo {author} {\bibfnamefont {D.~G.}\ \bibnamefont
  {Luchinsky}}, \bibinfo {author} {\bibfnamefont {R.~S.}\ \bibnamefont
  {Maier}}, \bibinfo {author} {\bibfnamefont {R.}~\bibnamefont {Mannella}},
  \bibinfo {author} {\bibfnamefont {P.~V.~E.}\ \bibnamefont {McClintock}}, \
  and\ \bibinfo {author} {\bibfnamefont {D.~L.}\ \bibnamefont {Stein}},\ }\href
  {\doibase 10.1103/PhysRevLett.82.1806} {\bibfield  {journal} {\bibinfo
  {journal} {Phys. Rev. Lett.}\ }\textbf {\bibinfo {volume} {82}},\ \bibinfo
  {pages} {1806} (\bibinfo {year} {1999}{\natexlab{b}})}\BibitemShut {NoStop}%
\bibitem [{\citenamefont {Feng}\ \emph {et~al.}(2014)\citenamefont {Feng},
  \citenamefont {Zhang},\ and\ \citenamefont {Wang}}]{Feng_2014}%
  \BibitemOpen
  \bibfield  {author} {\bibinfo {author} {\bibfnamefont {H.}~\bibnamefont
  {Feng}}, \bibinfo {author} {\bibfnamefont {K.}~\bibnamefont {Zhang}}, \ and\
  \bibinfo {author} {\bibfnamefont {J.}~\bibnamefont {Wang}},\ }\href {\doibase
  10.1039/C4SC00831F} {\bibfield  {journal} {\bibinfo  {journal} {Chem. Sci.}\
  }\textbf {\bibinfo {volume} {5}},\ \bibinfo {pages} {3761} (\bibinfo {year}
  {2014})}\BibitemShut {NoStop}%
\bibitem [{\citenamefont {Gottesman}\ \emph {et~al.}(2001)\citenamefont
  {Gottesman}, \citenamefont {Kitaev},\ and\ \citenamefont
  {Preskill}}]{Gottesman_2001}%
  \BibitemOpen
  \bibfield  {author} {\bibinfo {author} {\bibfnamefont {D.}~\bibnamefont
  {Gottesman}}, \bibinfo {author} {\bibfnamefont {A.}~\bibnamefont {Kitaev}}, \
  and\ \bibinfo {author} {\bibfnamefont {J.}~\bibnamefont {Preskill}},\ }\href
  {\doibase 10.1103/PhysRevA.64.012310} {\bibfield  {journal} {\bibinfo
  {journal} {Phys. Rev. A}\ }\textbf {\bibinfo {volume} {64}},\ \bibinfo
  {pages} {012310} (\bibinfo {year} {2001})}\BibitemShut {NoStop}%
\bibitem [{\citenamefont {Devoret}\ and\ \citenamefont
  {Schoelkopf}(2013)}]{Devoret_2013}%
  \BibitemOpen
  \bibfield  {author} {\bibinfo {author} {\bibfnamefont {M.~H.}\ \bibnamefont
  {Devoret}}\ and\ \bibinfo {author} {\bibfnamefont {R.~J.}\ \bibnamefont
  {Schoelkopf}},\ }\href {\doibase 10.1126/science.1231930} {\bibfield
  {journal} {\bibinfo  {journal} {Science}\ }\textbf {\bibinfo {volume}
  {339}},\ \bibinfo {pages} {1169} (\bibinfo {year} {2013})}\BibitemShut
  {NoStop}%
\bibitem [{\citenamefont {Landau}\ and\ \citenamefont
  {Lifshitz}(1976)}]{Landau_Lifshitz}%
  \BibitemOpen
  \bibfield  {author} {\bibinfo {author} {\bibfnamefont {L.}~\bibnamefont
  {Landau}}\ and\ \bibinfo {author} {\bibfnamefont {E.}~\bibnamefont
  {Lifshitz}},\ }\href@noop {} {\emph {\bibinfo {title} {Mechanics}}},\
  Butterworth-Heinemann\ (\bibinfo {year} {1976})\BibitemShut {NoStop}%
\bibitem [{\citenamefont {Košata}\ \emph {et~al.}(2022)\citenamefont
  {Košata}, \citenamefont {del Pino}, \citenamefont {Heugel},\ and\
  \citenamefont {Zilberberg}}]{kovsata2022harmonicbalance}%
  \BibitemOpen
  \bibfield  {author} {\bibinfo {author} {\bibfnamefont {J.}~\bibnamefont
  {Košata}}, \bibinfo {author} {\bibfnamefont {J.}~\bibnamefont {del Pino}},
  \bibinfo {author} {\bibfnamefont {T.~L.}\ \bibnamefont {Heugel}}, \ and\
  \bibinfo {author} {\bibfnamefont {O.}~\bibnamefont {Zilberberg}},\ }\href
  {\doibase 10.21468/SciPostPhysCodeb.6} {\bibfield  {journal} {\bibinfo
  {journal} {SciPost Phys. Codebases}\ ,\ \bibinfo {pages} {6}} (\bibinfo
  {year} {2022})}\BibitemShut {NoStop}%
\end{thebibliography}

%

\end{document}